\documentclass[twocolumn,12pt]{article}
\usepackage[utf8]{inputenc}
\usepackage[OT4]{fontenc}
\usepackage{amsmath}
\usepackage{amssymb}
\usepackage{graphicx}
\usepackage{multirow}
\usepackage{longtable,lscape}
\usepackage{diagbox}
\usepackage{subfig}
\usepackage{wasysym}

\begin{document}
\title{Communities and classes in symmetric fractals}
\author{Małgorzata J. Krawczyk\\
Faculty of Physics and Applied Computer Science,\\
AGH University of Science and Technology,\\
al. Mickiewicza 30, 30-059 Cracow, Poland\\
e-mail: gos@fatcat.ftj.agh.edu.pl}

\maketitle

\begin{abstract}
Two aspects of fractal networks are considered: the community structure and the class structure, where classes of nodes appear as a consequence of a local symmetry of nodes. 
The analysed systems are the networks constructed for two selected symmetric fractals: the Sierpinski triangle and the Koch curve. Communities are searched for by means of a 
set of differential equations. Overlapping nodes which belong to two different communities are identified by adding some noise to the initial connectivity matrix. Then, a node 
can be characterized by a spectrum of probabilities of belonging to different communities. Our main goal is that the overlapping nodes with the same spectra belong to the same class.
\end{abstract}

\section{Introduction}
Analysis of properties of networks may regard different aspects. One of them is the identification of communities; it is equivalent to a division of the network into sub-networks, 
sometimes partially overlapping. An informal definition says that a community is a set of nodes which are more densely connected to each other than to remaining nodes of the 
system \cite{comm1, comm2}. An exhaustive review of the methods of identification of communities has been presented in \cite{comm3}. Here we note the well-known problem which 
arises when overlapping nodes are attached with the same strength to more than one community \cite{over}. Another important problem is that it is difficult to indicate communities 
properly when communities differ significantly in their sizes \cite{comm2}. Further, sometimes even though communities are well-defined, the link density is so homogeneous that it 
does not allow for their proper reconstruction of the communities by used algorithms \cite{comm4}.  \\

Analysis of networks can regard also local aspects of the network structure, as degree of nodes and their neighbours. In networks with some symmetry, these local properties may be 
the same for different nodes. This observation allows us to group nodes into classes which differentiate these local properties \cite{class}. Large size of networks is often an 
obstacle in simulations. The classification of nodes allows for an effective reduction of the system size which may preserve an information of system properties \cite{class}.\\

The main issue of the paper is what we can infer about the class structure from the results of the analysis of the community structure.\\ 

The paper is organised as follows: At first,the analysed networks are presented together with the methods of communities and classes identification. The following 
Section \ref{results} is devoted to the presentation of the results obtained from both these procedures. The last section contains our concluding remarks.

\section{The analysed networks}
As examples of networks which display both overlapping and symmetry, we use the networks constructed on bases of two instances of the Lindenmayer systems (L-systems) \cite{lsys} -- the Sierpinski triangle and the Koch curve. In the first case, apexes of triangles are identified with network's nodes (while we take into account that most of apexes belong to more than one triangle), and sides of triangles are identified with network's edges. In the case of the Koch curve, the network nodes are identified with curve intersects. In the latter case, we observe not only symmetry and overlapping communities but also differences of community sizes.

\subsection{Communities}
To identify communities we use the method proposed by us some years ago \cite{comm}. The idea of the method is that a link weight between two nodes increases if these nodes have a common neighbour. According to the proposed set of differential equations, the connectivity matrix evolve, and weights of some edges increase while others decrease. The evolution equation is of the form:
\[
\dfrac{dA_{ij}}{dt}=G(A_{ij})\sum\limits_{k\ne i,j}(A_{ij}A_{kj}-\beta)
\label{dif}
\]
where $A_{ij}$ - an element of the connectivity matrix, $G(x)=\Theta(x)\Theta(1-x)$, and $\beta$ is the model parameter, which specifies a threshold between meaningful and negligible
values of the product of two links weights. Under these equations, the time evolution of the method is fully deterministic, and therefore free from an inherent noise of Monte Carlo methods. 
The evolution process leads to a subsequent division of the nodes into communities. The accepted division is the one with the highest modularity value \cite{mod1, mod2}.\\

Here the connectivity matrix constructed for the analysed networks consists of 1's and 0's. If any two nodes are connected an appropriate matrix element is equal to $1$, otherwise it is $0$. Application of our method to such a network gives only communities which are well defined. In the case of the networks analysed here, one can expect that overlapping nodes will be identified as one-node communities. However, to be able indicate connections of those nodes with other nodes in the network we slightly disturb the connectivity matrix. Namely, we add a random number to each 0 and we subtract a random number from each 1. If the amplitude of random numbers is small enough, such a modification - for sufficiently big statistics - does not destroy the network structure. Yet, an application of the method allows to divide the network into communities which contain overlapping nodes.\\ As our procedure involves a disturbance of the connectivity matrix by some random number it is necessary to indicate the amplitude of 
the used noise. This means that now we have two parameters: just mentioned amplitude and the parameter $\beta$ in the community identification procedure. The values of those parameters were found experimentally for each analysed network to obtain the division into communities which guarantees a maximal value of the modularity.

\subsection{Classes}
The idea of the classes is based on the symmetry observed in networks, which causes that it is possible to indicate groups of states which are equivalent. The symmetry here means that 
some nodes have the same number of neighbours which belong to the same classes. This observation allows for the reduction of the system size, as the whole network can be represented 
in the equivalent way as the network of classes \cite{class, class1, class2}.  The procedure of classes identification takes into account not only the number of neighbours of each 
node, but also the weights of connections. In particular, in some cases weights of edges between two nodes in both directions (e.g. $a\rightarrow b$ and $a\leftarrow b$) may be different. Because of that the full procedure takes into account the lists of in- and out-neighbours of each node. In the case of networks analysed here, the procedure may be simplified as networks are unweighted and undirected. The noise, remarked in the preceding section, is applied only in the 
search of communities, and not classes. Details of the procedure of classes identification have been presented in \cite{class2}. We present here only the main features of the method. 
The procedure of class identification is performed as follows: At first we check degrees of the nodes of the network. Then, we assign to each node which has a given degree a unique
symbol (class identifier). Subsequently we make lists of neighbours for each node, where we replace the node number with its class symbol. If the lists of neighbours for all nodes which
have assigned the same symbol are not the same, we have to introduce additional distinction between nodes with the same degree. The classification is repeated until unique lists of 
neighbours for different class symbols are obtained.

\section{Results}
\label{results}
\subsection{The Sierpinski triangle - the communities}
\begin{figure}[!hptb]
\begin{center}
\includegraphics[width=.7\columnwidth, angle=0]{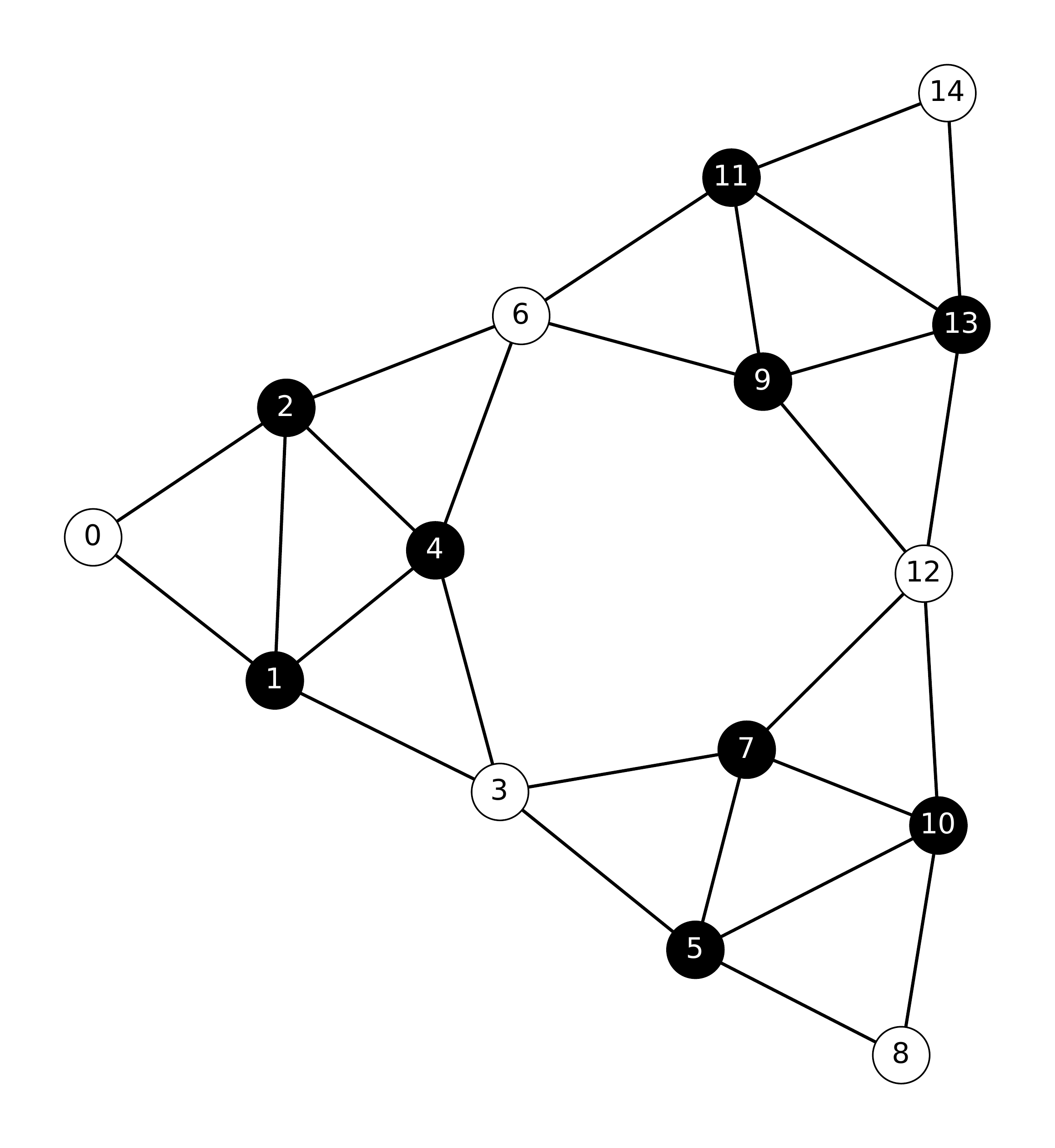}
\caption{The Sierpinski triangle network with indication of communities for $N=15$. Here we have $3$ communities of $3$ nodes each (black nodes) and $6$ communities of $1$ node each (white nodes)}
\label{fig:comm15}
\end{center}
\end{figure}

\begin{figure}[!hptb]
\begin{center}
\subfloat[Node number $0$, analogous histograms are obtained for nodes $8$ and $14$ with connection to the community $2$ and $3$, respectively.]{\includegraphics[width=.7\columnwidth, angle=0]{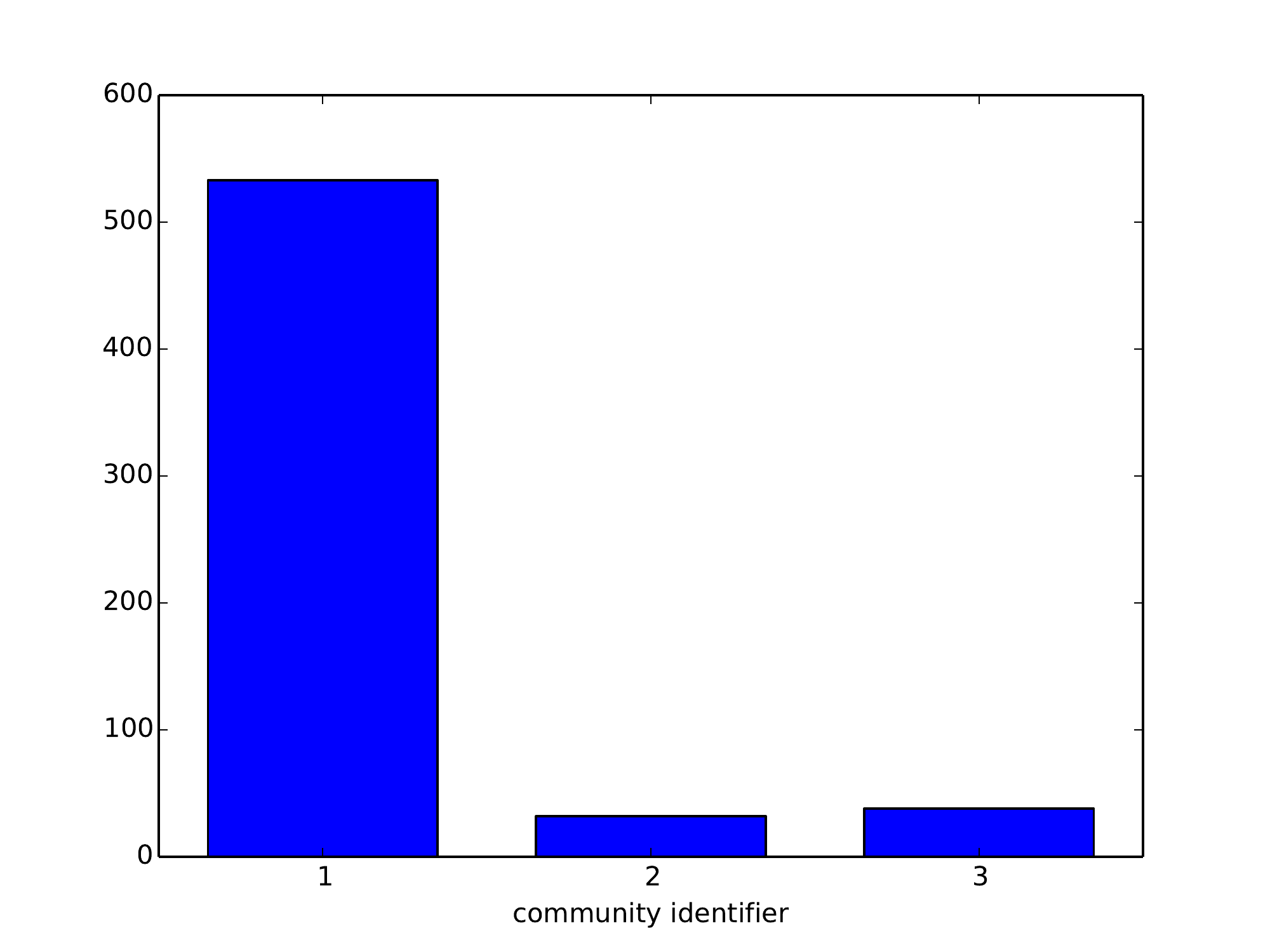}\label{fig:hist15a}}\\
\subfloat[Node number $3$, analogous histograms are obtained for nodes $6$ and $12$ with connection to the community $1$ and $3$, and $2$ and $3$, respectively.]{\includegraphics[width=.7\columnwidth, angle=0]{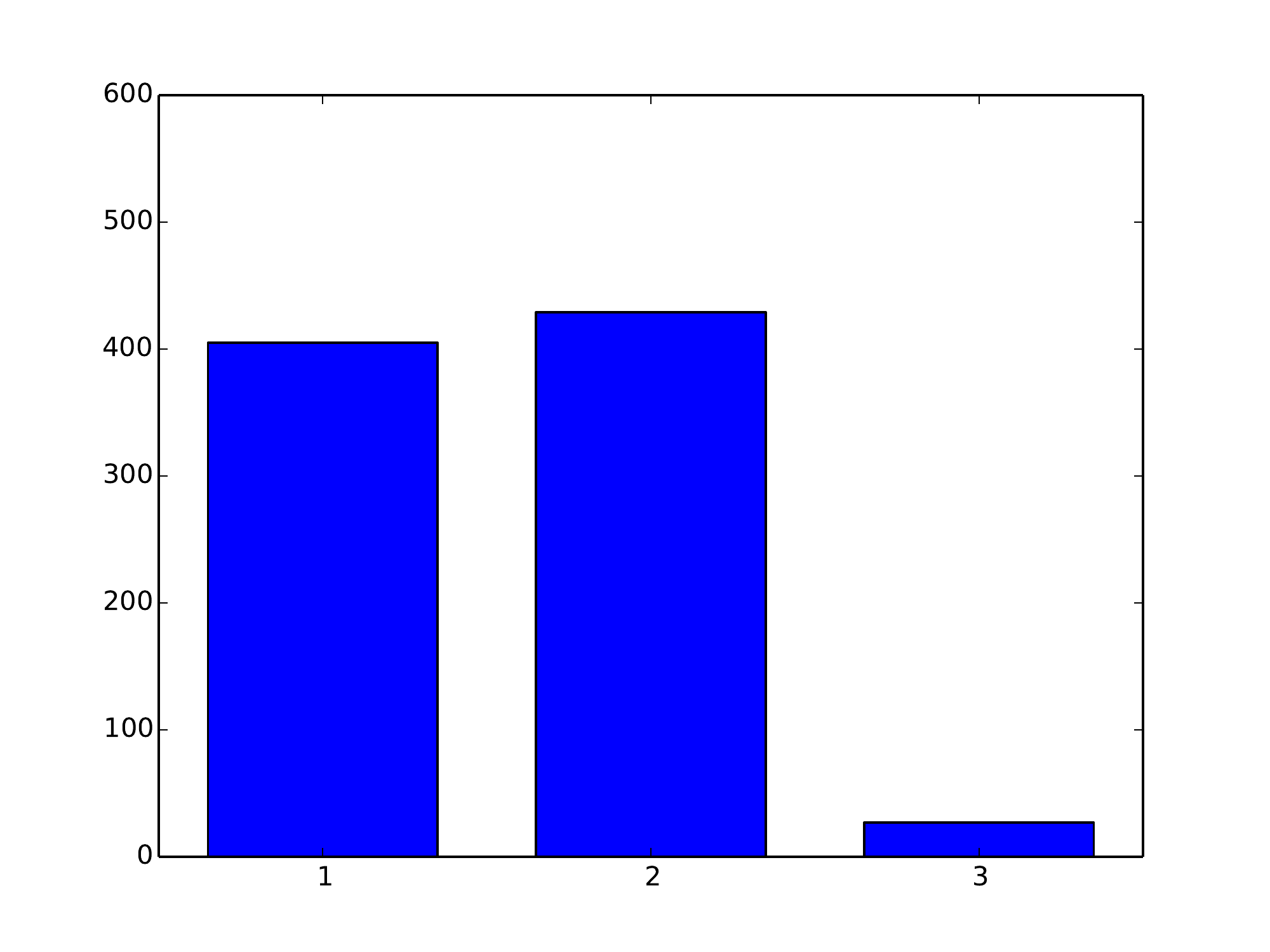}\label{fig:hist15b}}
\caption{Histograms of the frequency of the connection of a given node with original communities of the size different than one for the Sierpinski triangle network for $N=15$.}
\label{fig:hist15}
\end{center}
\end{figure}

\textit{$\bullet$ $N=15$, no noise:} In the case of the Sierpinski triangle network consisting of $15$ nodes, the community identification method applied to the original connectivity matrix leads to the division of the network into $3$ communities of size $3$ each (community $1$ - nodes $1$, $2$ and $4$, community $2$ - nodes $5$, $7$ and $10$, and community $3$ - nodes $9$, $11$ and $13$) and $6$ communities of $1$ node each. The result is presented in Fig.\ref{fig:comm15}. For this division, the modularity is equal to $0.12$ for $\beta=0.4$. Here we see that the nodes which form one-element communities are of two types: the first type for nodes number $0$, $8$ and $14$ and the second for nodes $3$, $6$ and $12$.\\
\textit{$\bullet$ $N=15$, noise:} To indicate the connections between communities we add a noise of amplitude $0.01$ to the connectivity matrix elements and we record the obtained communities for independent $1000$ realizations of the noise.  The maximal obtained modularity is $0.19$, for $\beta=0.15$ and the division of the network into $3$ communities. The obtained result can be presented in the form of histograms of the frequency of a given node being connected to a given original community of the size different than one. Examples of the obtained histograms are presented in Fig.\ref{fig:hist15}.  As we see, two different situations are observed for two mentioned types of nodes which originally form one node communities. For the first type (0, 8, 14), one clear maximum is observed, and for the second type (3, 6, 12), we observe two values significantly higher than the third one. This result indicates that the nodes of the second type can be equally likely assigned to two different communities. In this case we deal with the 
overlapping.\\

\begin{figure}[!hptb]
\begin{center}
\includegraphics[width=.7\columnwidth, angle=0]{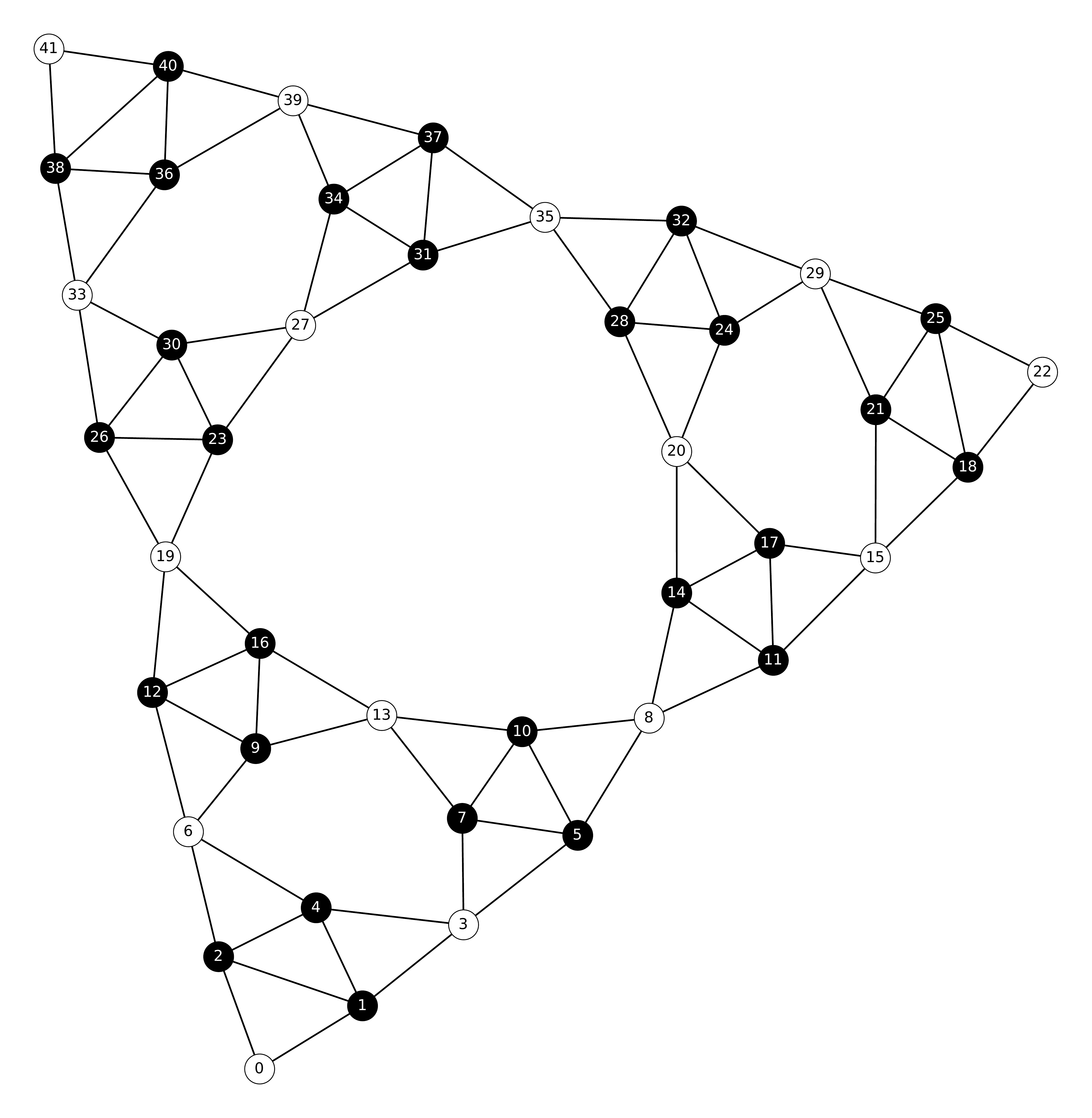}
\caption{The Sierpinski triangle network with indication of communities for $N=42$, here we have $9$ communities of $3$ nodes each (black nodes) and $15$ communities of $1$ node each (white nodes).}
\label{fig:comm42}
\end{center}
\end{figure}

\begin{figure*}[!hptb]
\begin{center}
\subfloat[Node number $0$. Analogous histograms are obtained for nodes $22$ and $41$.]{\includegraphics[width=.7\columnwidth, angle=0]{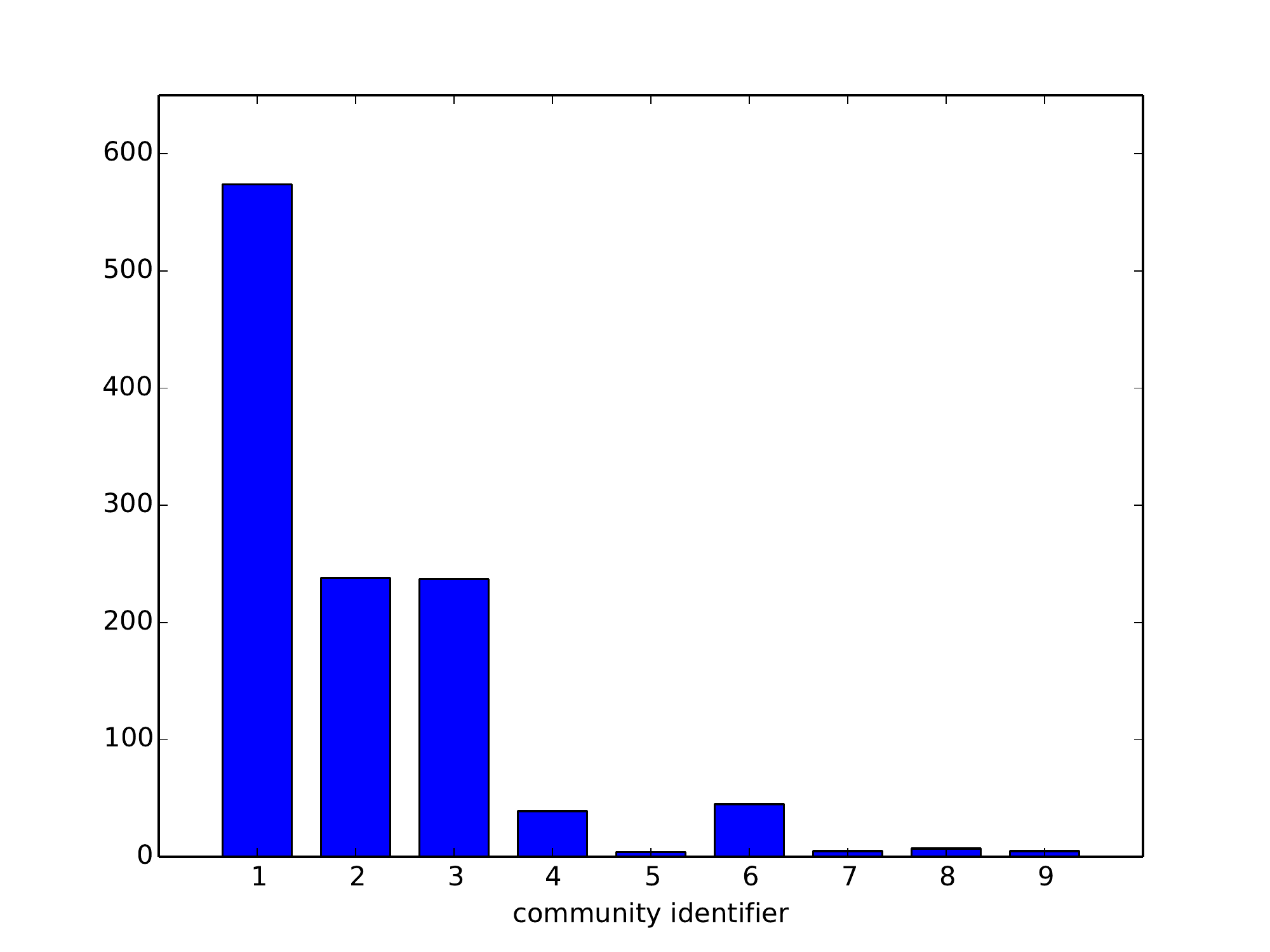}\label{fig:hist42a}}
\hspace{1cm}
\subfloat[Node number $3$. Analogous histograms are obtained for nodes $6$, $15$, $29$, $33$ and $39$.]{\includegraphics[width=.7\columnwidth, angle=0]{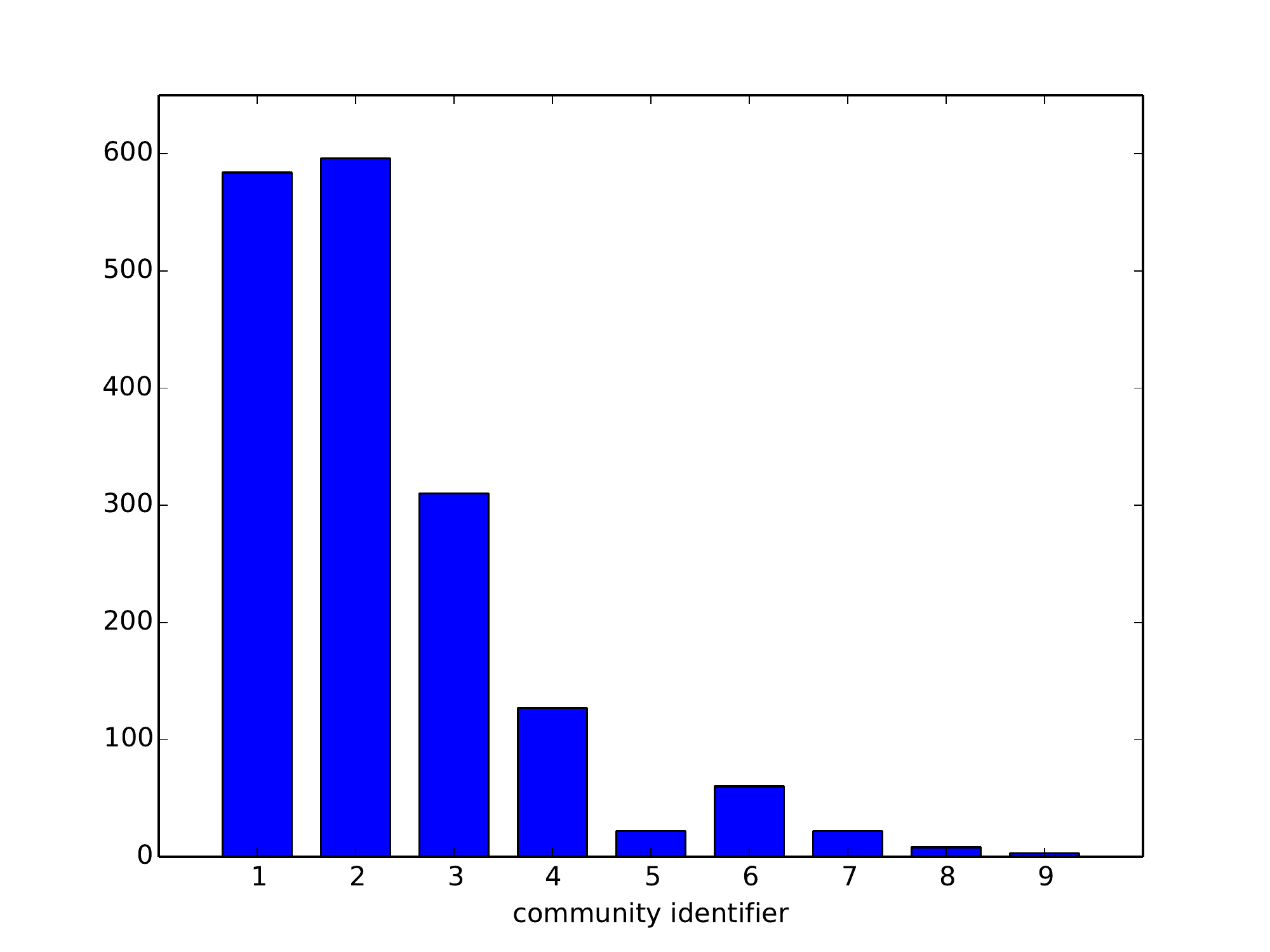}\label{fig:hist42b}}\\
\subfloat[Node number $8$. Analogous histograms are obtained for nodes $19$ and $35$.]{\includegraphics[width=.7\columnwidth, angle=0]{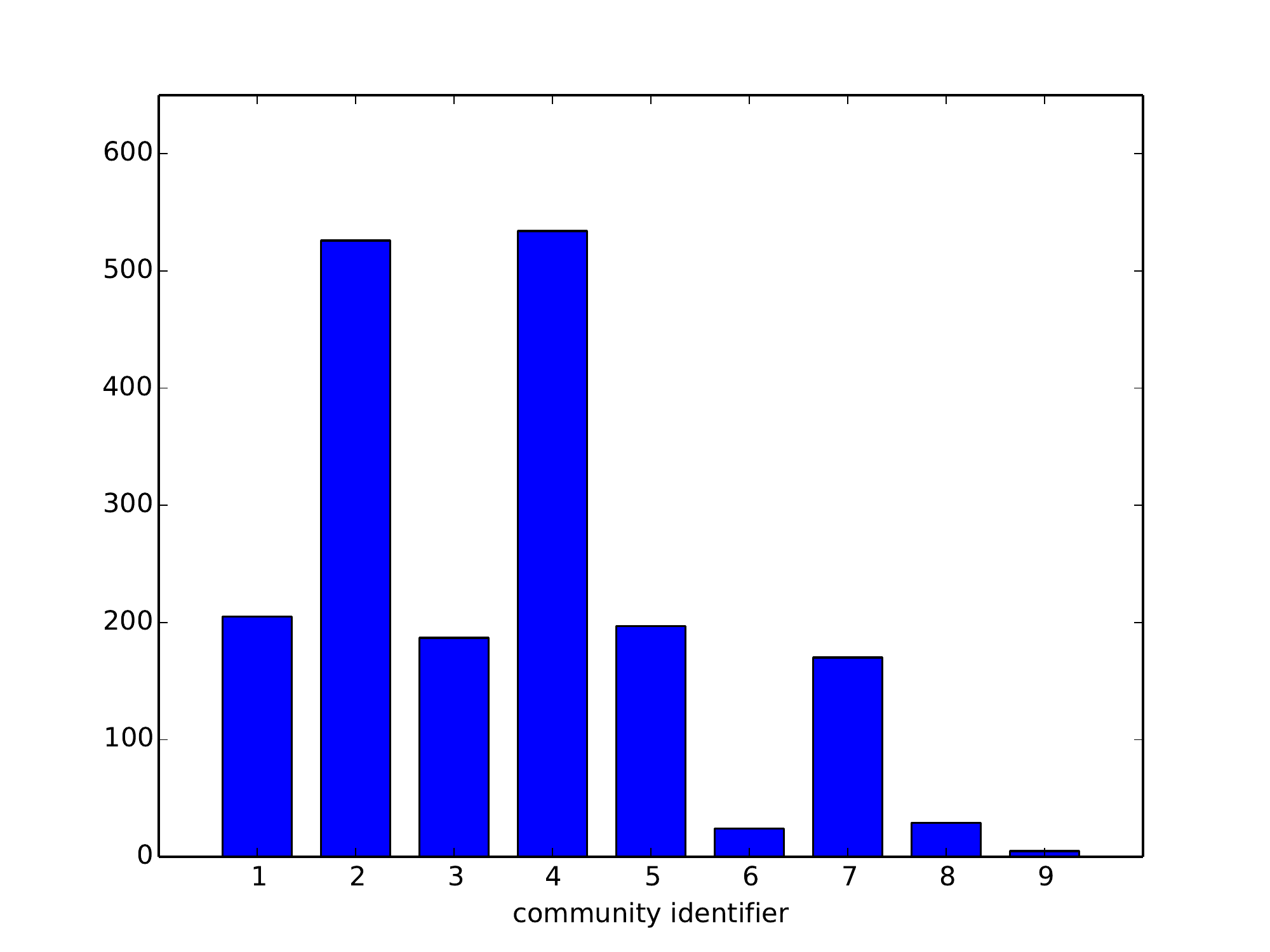}\label{fig:hist42c}}
\hspace{1cm}
\subfloat[Node number $13$. Analogous histograms are obtained for nodes $20$ and $27$.]{\includegraphics[width=.7\columnwidth, angle=0]{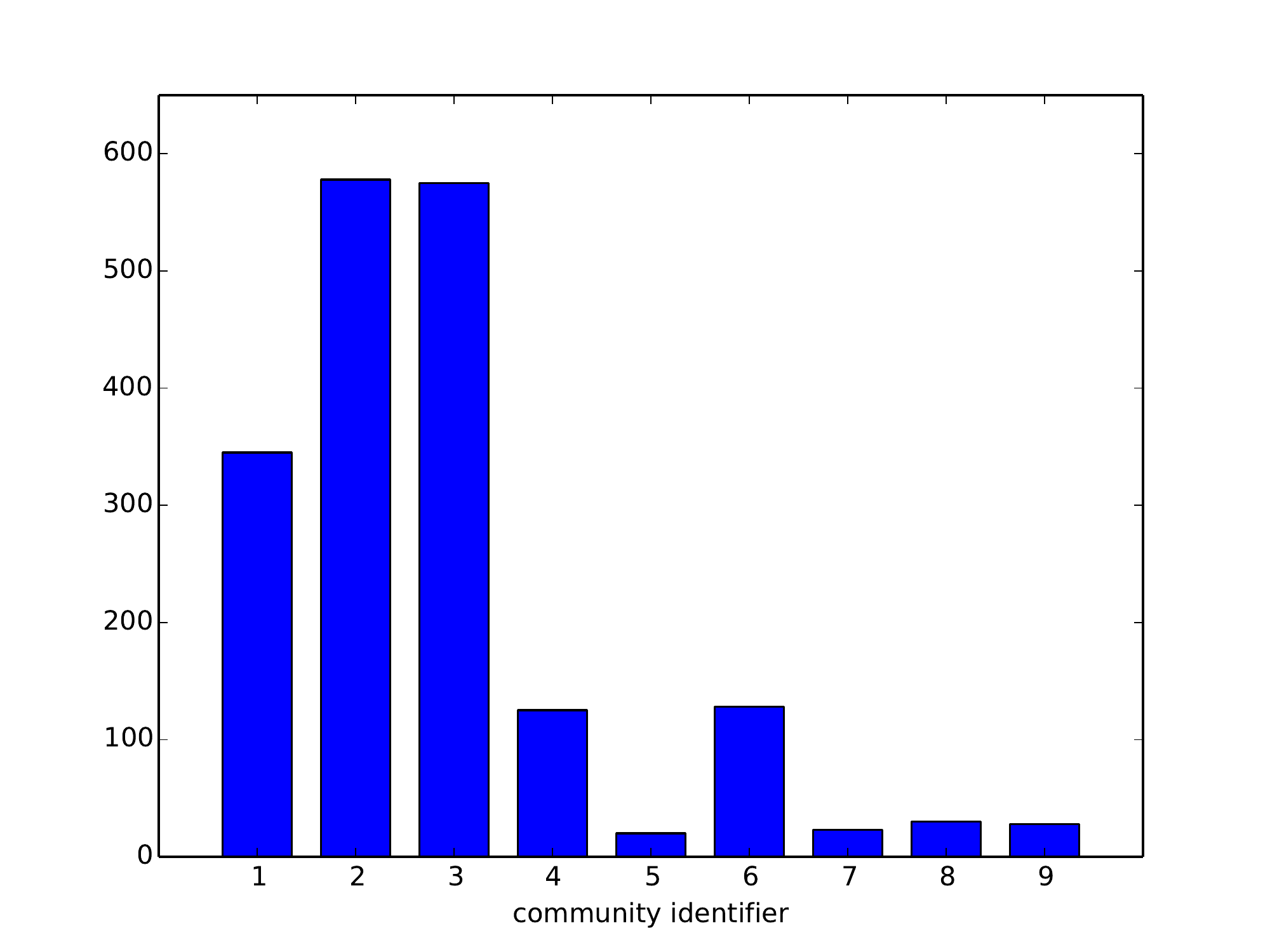}\label{fig:hist42d}}
\caption{Histograms of the frequency of the connections of a given node with original communities of the size different than one for the Sierpinski triangle network for $N=42$.}
\label{fig:hist42}
\end{center}
\end{figure*}

\noindent
\textit{$\bullet$ N=42, no noise:} For a larger network of $N=42$ we obtain $9$ communities of $3$ nodes each and $15$ communities of $1$ node each, for original connectivity matrix (see Fig.\ref{fig:comm42}).  For this network the modularity equals $0.15$ for $\beta=0.4$.\\
\textit{$\bullet$ N=42, noise:} In this case, for the connectivity matrix with noise of amplitude $0.001$ and for $\beta=0.05$, four types of nodes classified to one node communities are observed, which is reflected in the histograms profiles. The optimal division is obtained for modularity equal to $0.26$ and division in $9$ communities. The first type form nodes number $0$, $22$ and $41$, the second type - nodes $3$, $6$, $15$, $29$, $33$ and $39$, the third type nodes $8$, $19$, $35$, and the last one -- nodes $13$, $20$ and $27$ (see Fig.\ref{fig:hist42}). The histograms obtained in this case allow not only to indicate the communities to which a given node is connected to, but also reveal further connections between them. A similar frequency indicate the same "distance" to a given community.

In Fig.\ref{fig:hist42a} we see one maximum which indicate the community with which the node $0$ is directly connected to. Two subsequent bars with approximately equal height indicate the communities which are next nearest neighbouring communities. In Figs.\ref{fig:hist42b} and \ref{fig:hist42d} we see two bars approximately equal in size, which indicate nearest communities to the node number $3$. The lower bar is related to the community which is connected with both just mentioned communities, and so on. Alike, in Fig.\ref{fig:hist42c} there are two direct neighbouring communities, and four next neighbouring communities.\\


\subsection{The Koch curve - the communities}
\begin{figure}[!hptb]
\begin{center}
\includegraphics[width=.7\columnwidth, angle=0]{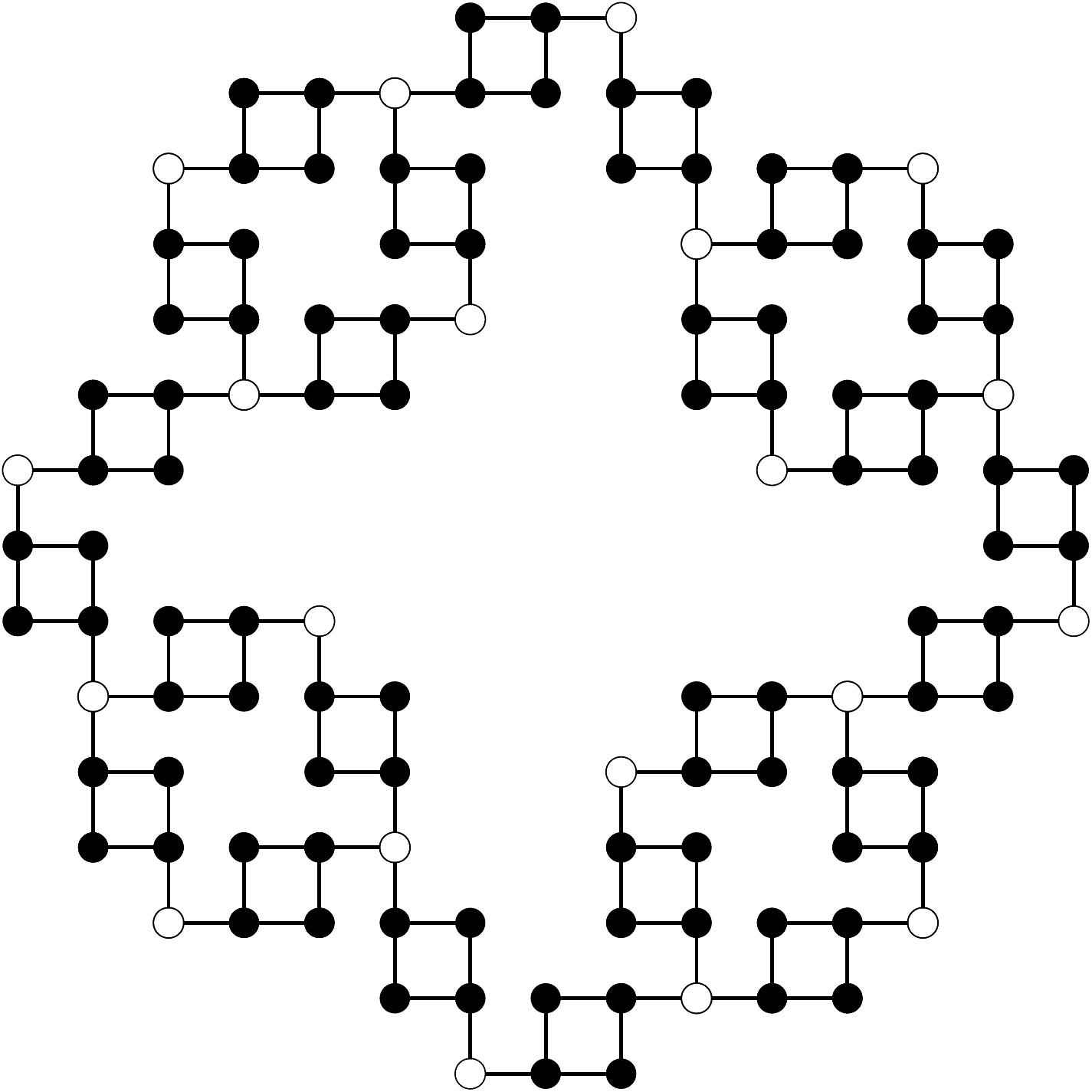}
\caption{The Koch curve network with indication of communities for $N=116$, here we have $24$ communities of $4$ nodes each (black nodes) and $20$ communities of $1$ node each (white nodes).}
\label{fig:comm116}
\end{center}
\end{figure}

\begin{figure}[!hptb]
\begin{center}
\subfloat[First type, analogous histograms are obtained for $7$ similar nodes.]{\includegraphics[width=.7\columnwidth, angle=0]{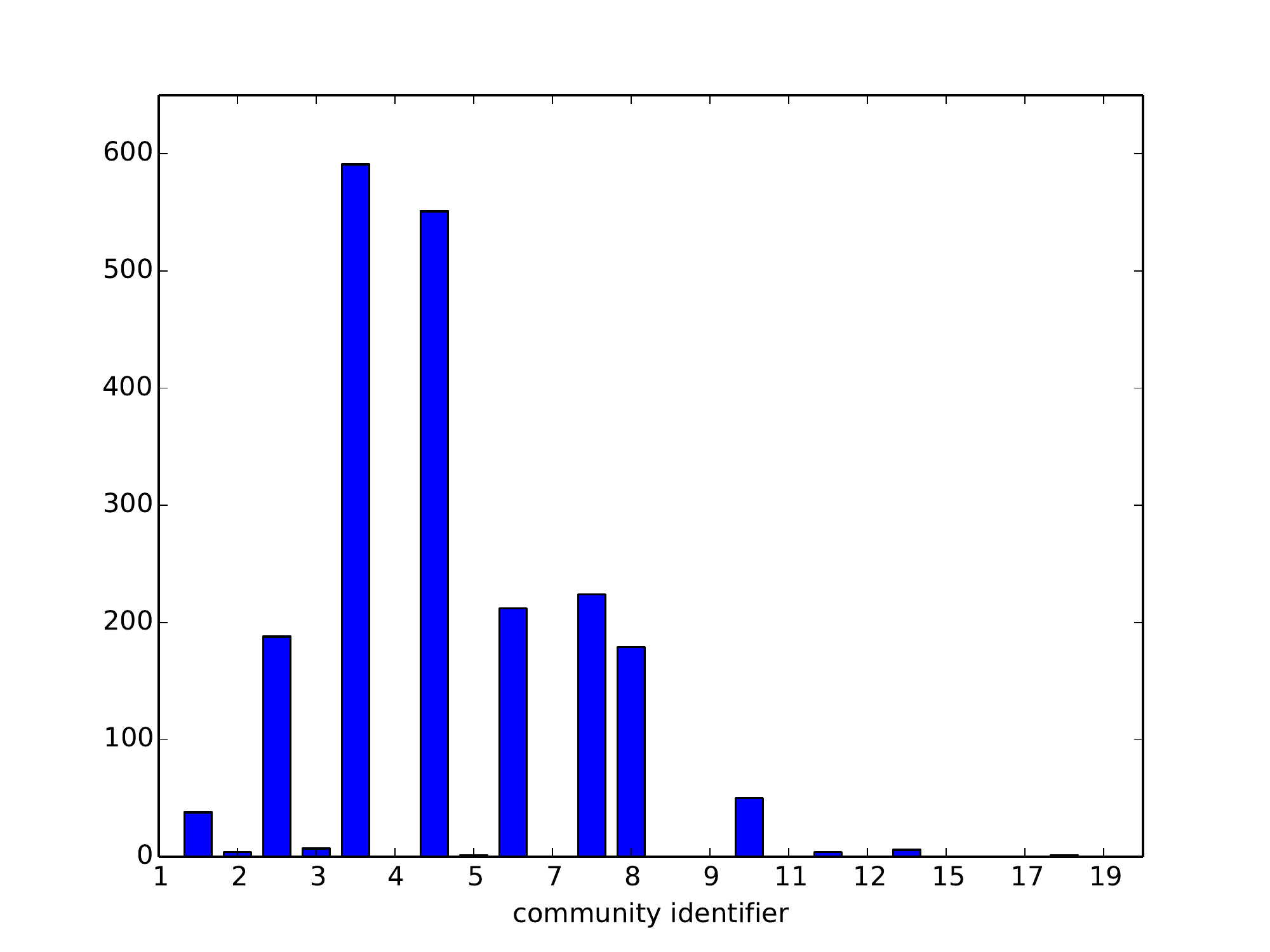}\label{fig:hist116a}}\\
\subfloat[Second type, analogous histograms are obtained for $3$ similar nodes.]{\includegraphics[width=.7\columnwidth, angle=0]{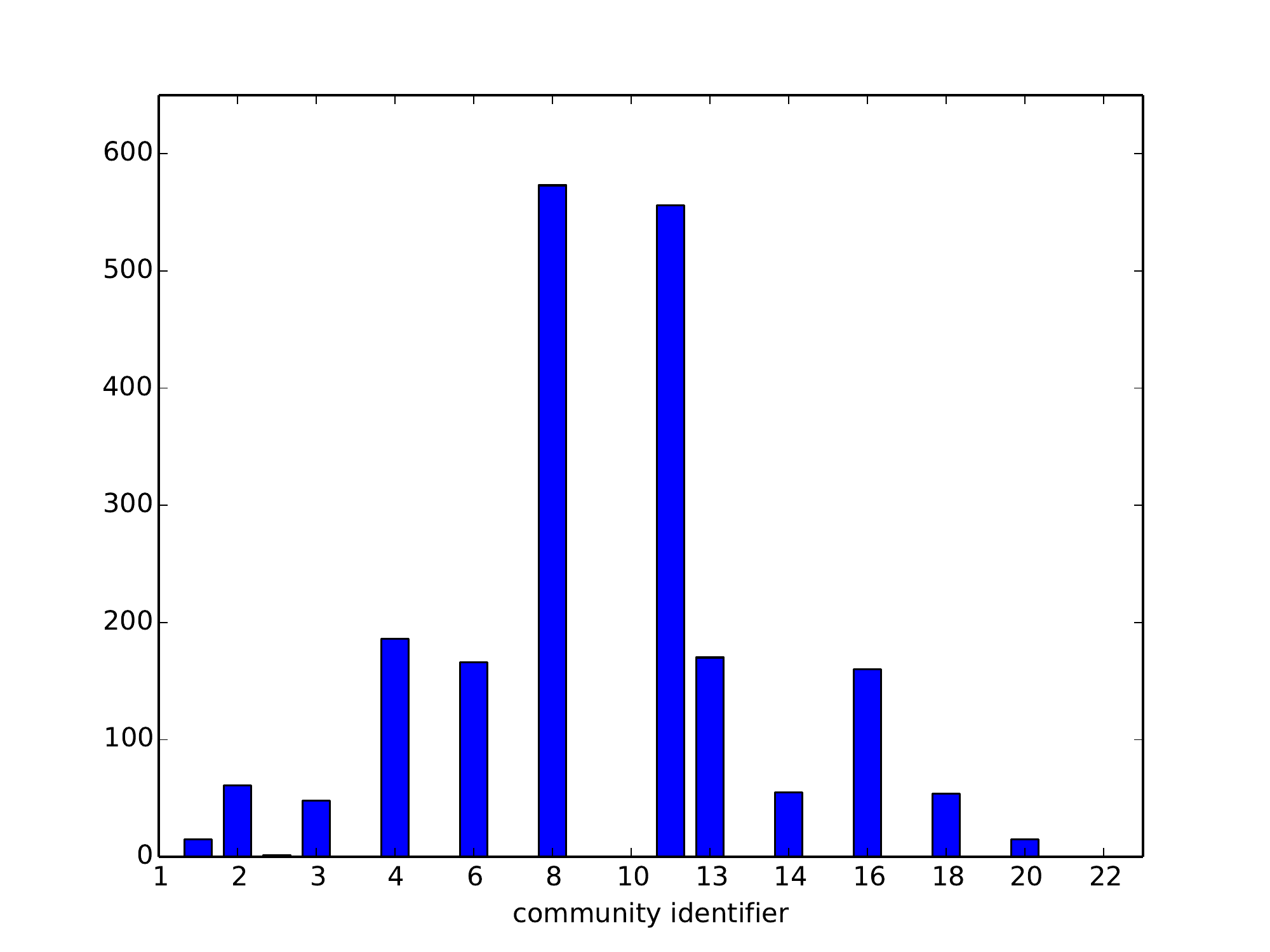}\label{fig:hist116b}}\\
\subfloat[Third type, analogous histograms are obtained for $7$ similar nodes.]{\includegraphics[width=.7\columnwidth, angle=0]{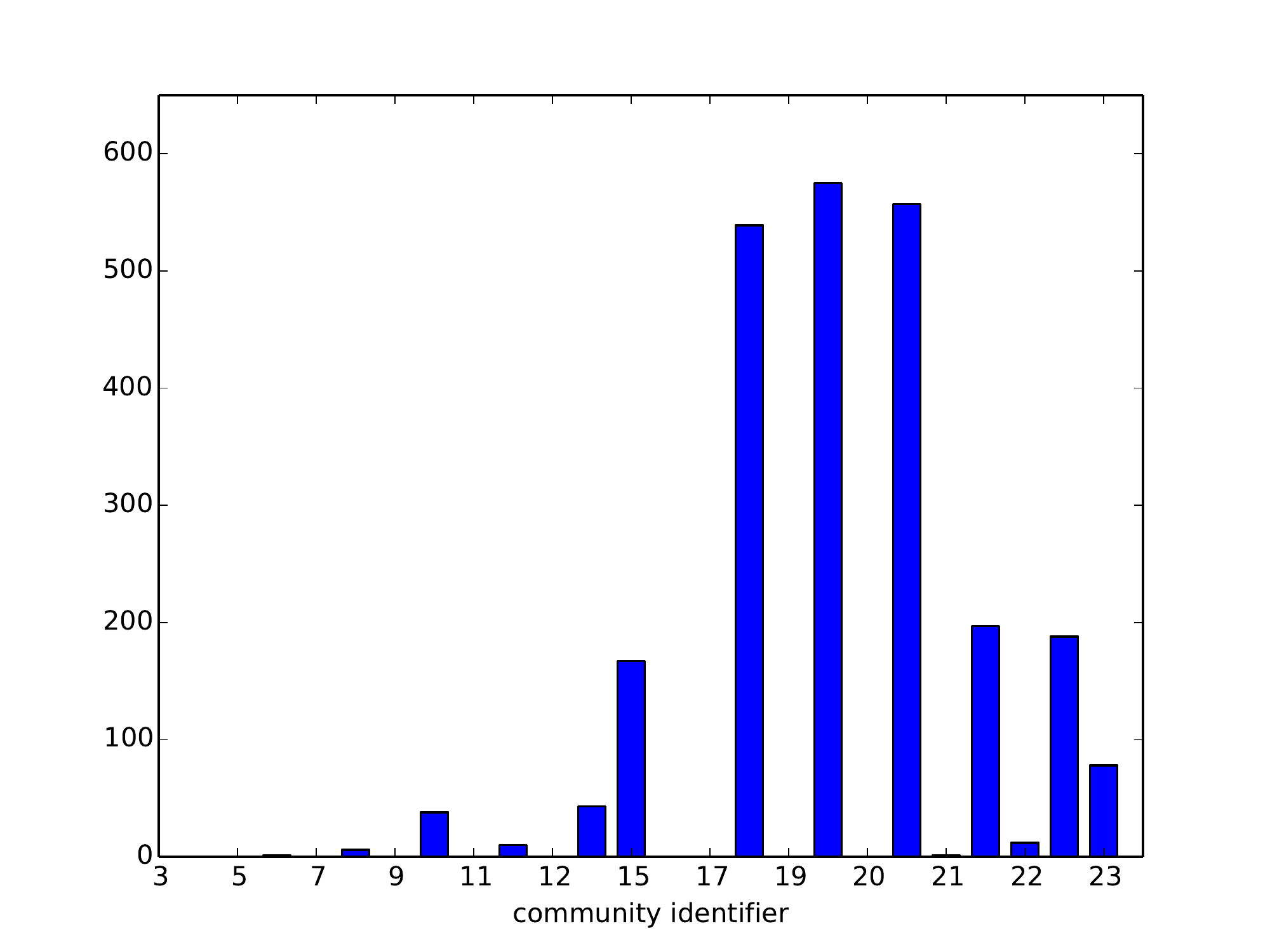}\label{fig:hist116c}}
\caption{Histograms of the frequency of the connection of a given node with original communities of the size different than one for the Koch curve for $N=116$.}
\label{fig:hist116}
\end{center}
\end{figure}

\textit{$\bullet$ N=116, no noise:} In the case of the Koch curve network of size $N=116$ the original system is divided into $24$ communities of size $4$ each and $20$ communities of $1$ node each (Fig.\ref{fig:comm116}). Such a division was obtained for $\beta=0.01$ and leads to modularity equal to $0.32$. The same usage of noise allows to indicate the overlapping nodes.  Typical histograms obtained for this network are presented in Fig.\ref{fig:hist116}.

\textit{$\bullet$ N=116, noise:} The maximal value of modularity for noisy network the modularity equals $0.42$ and is obtained for noise amplitude equal to $0.001$ and $\beta=0.05$. The number of communities for those parameters equals $24$. Here we observe three profiles of histograms, which reflect three types of white nodes in Fig.\ref{fig:comm116}. We see there two different types of nodes of degree $2$ and one node type of nodes of degree $3$. In the histograms it manifests in $2$ or $3$ bars higher than other bars (Fig.\ref{fig:hist116}). Also a distinction between nodes of degree $2$ is observable in the histogram profiles (see Figs.\ref{fig:hist116a} and \ref{fig:hist116b}).

\begin{figure}[!hptb]
\begin{center}
\includegraphics[width=.7\columnwidth, angle=0]{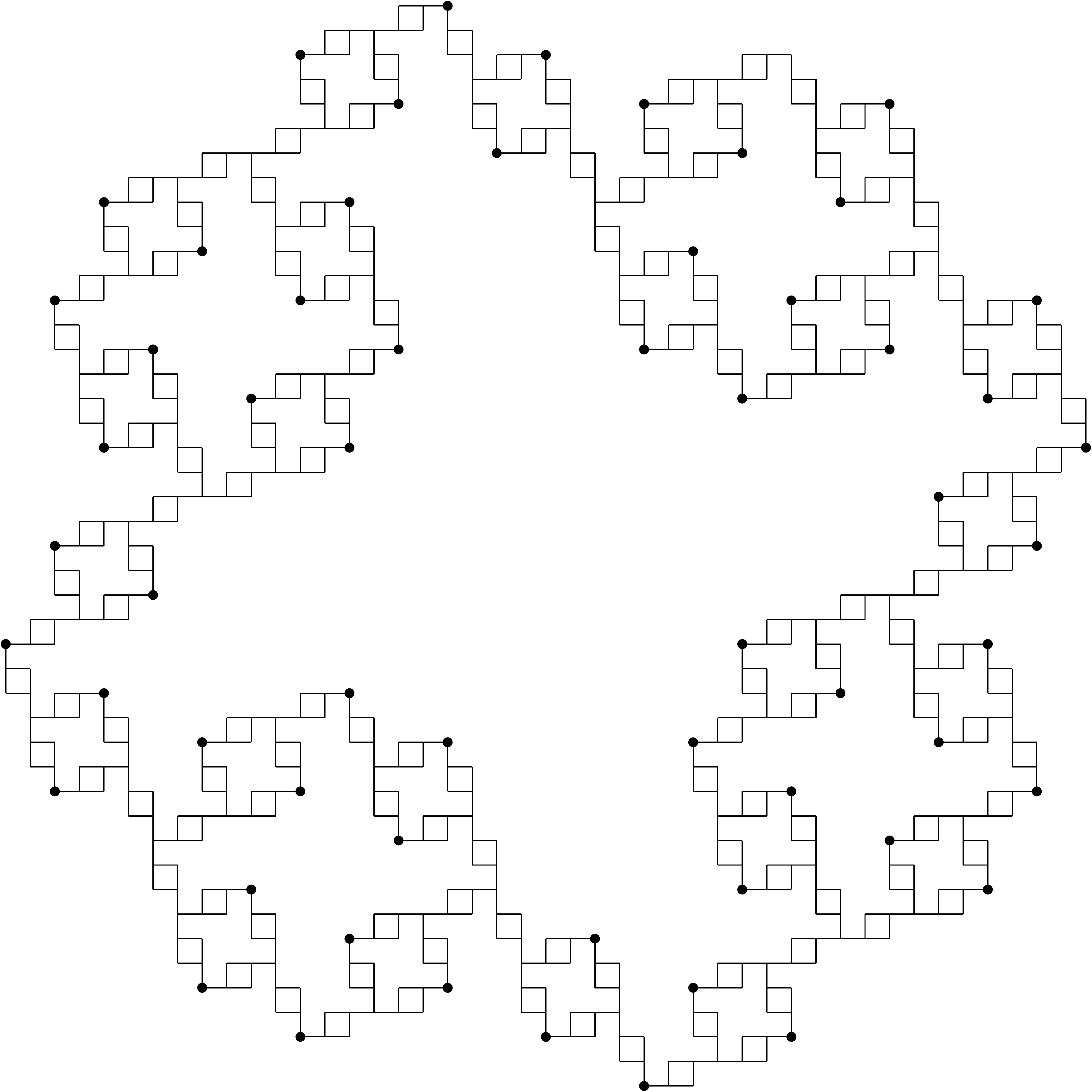}
\caption{The Koch curve network for $N=692$, black dots indicate nodes classified to singular communities.}
\label{fig:comm692}
\end{center}
\end{figure}

\textit{$\bullet$ N=692, no noise:} For a larger system of size $N=692$ nodes, $24$ communities of size $13$, $8$ communities of size $40$, and $60$ nodes of size $1$ are obtained. The division is presented in Fig.\ref{fig:comm692}, where black dots indicate nodes classified to singular communities. Such a division - obtained for $\beta=0.001$ - leads to modularity value $0.41$.

\textit{$\bullet$ N=692, noise:} Noise amplitude $0.001$ and $\beta=0.08$ leads to the division of the network into $32$ communities with modularity equal to $0.36$ ($220$ repetitions).

\begin{figure*}[!hptb]
\begin{center}
\subfloat[First type, analogous histograms are obtained for $3$ similar nodes.]{\includegraphics[width=.7\columnwidth, angle=0]{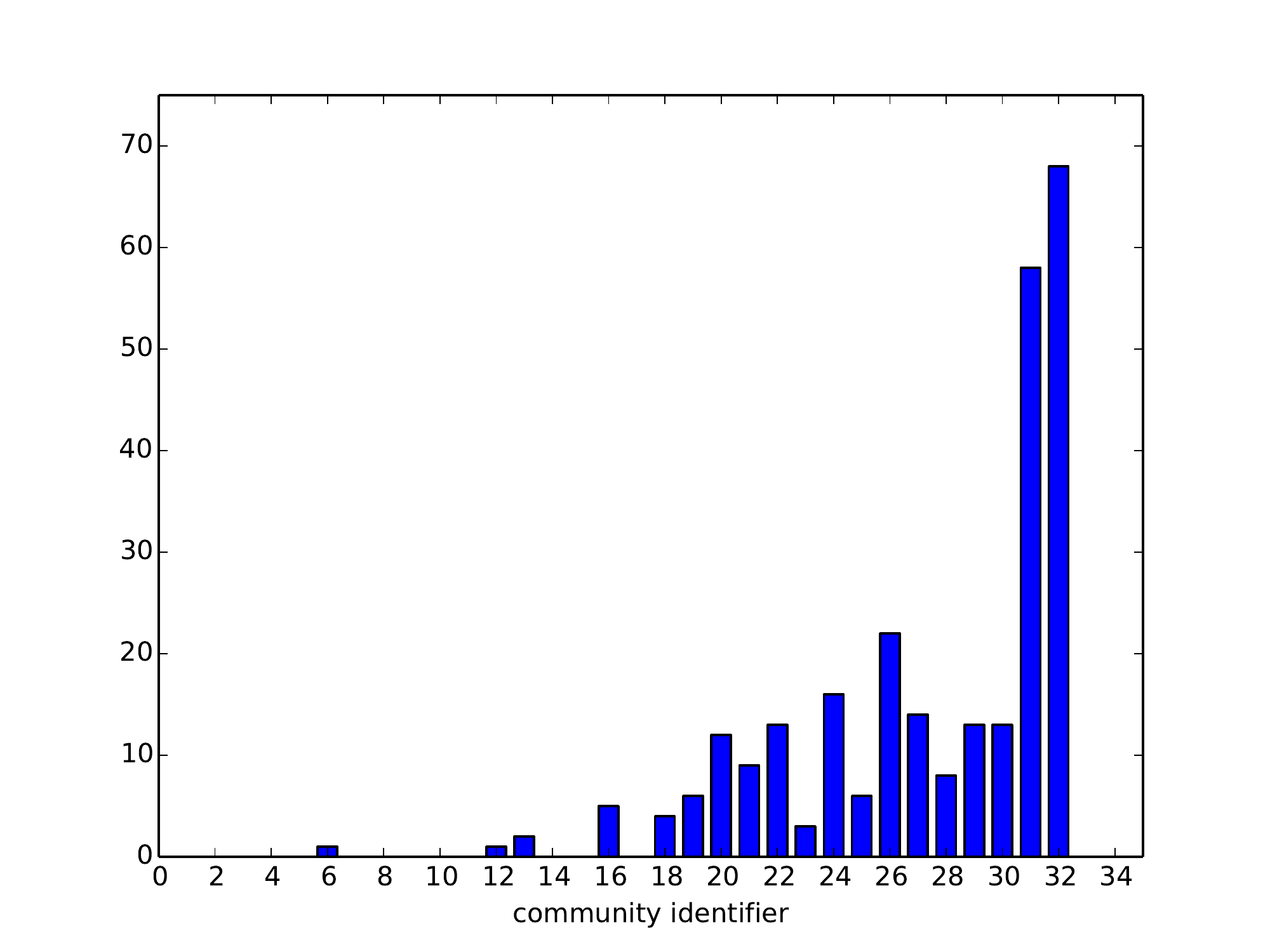}\label{fig:hist692a}}
\subfloat[Second type, analogous histograms are obtained for $7$ similar nodes.]{\includegraphics[width=.7\columnwidth, angle=0]{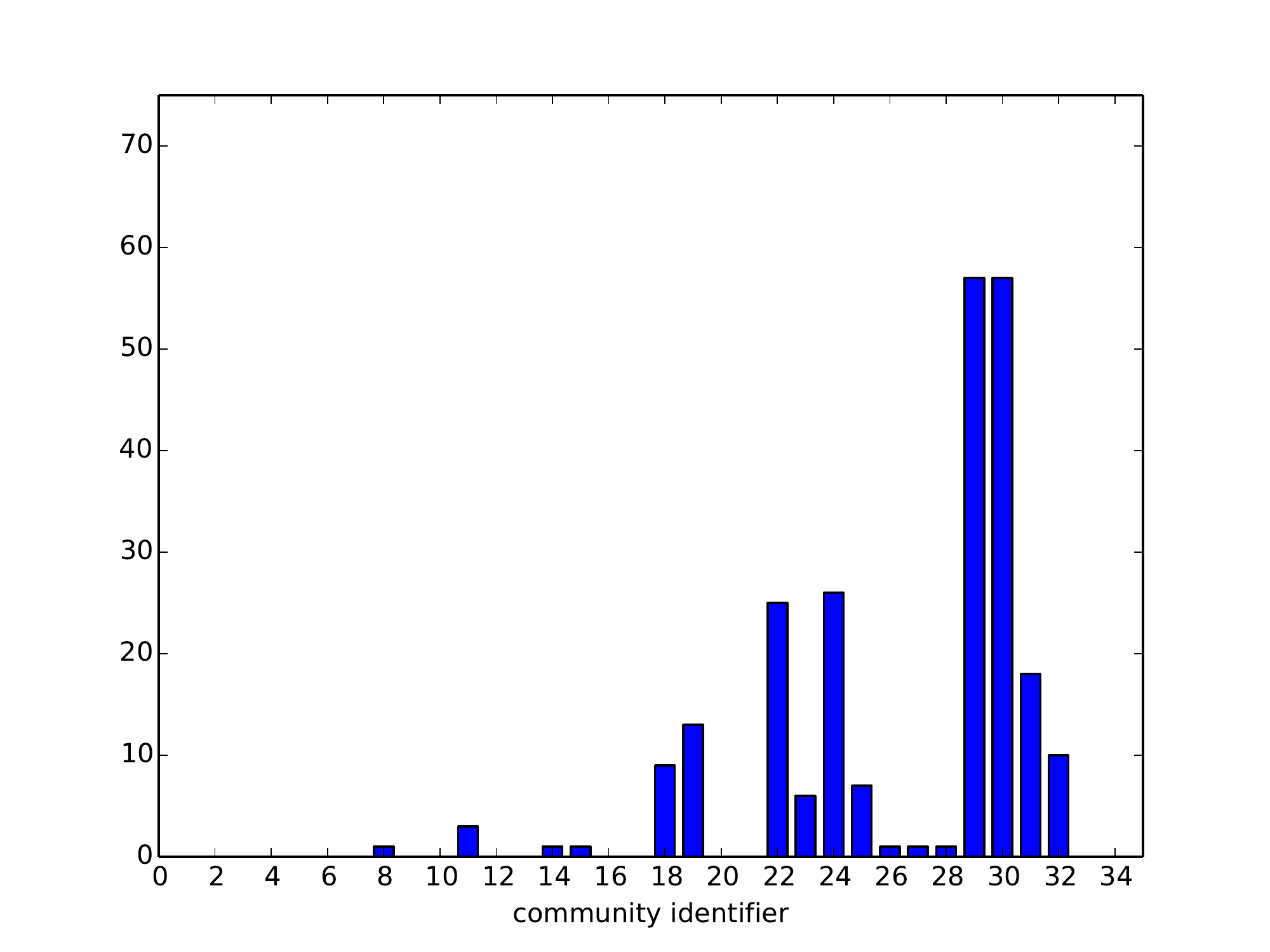}\label{fig:hist692b}}\\
\subfloat[Third type, analogous histograms are obtained for $15$ similar nodes.]{\includegraphics[width=.7\columnwidth, angle=0]{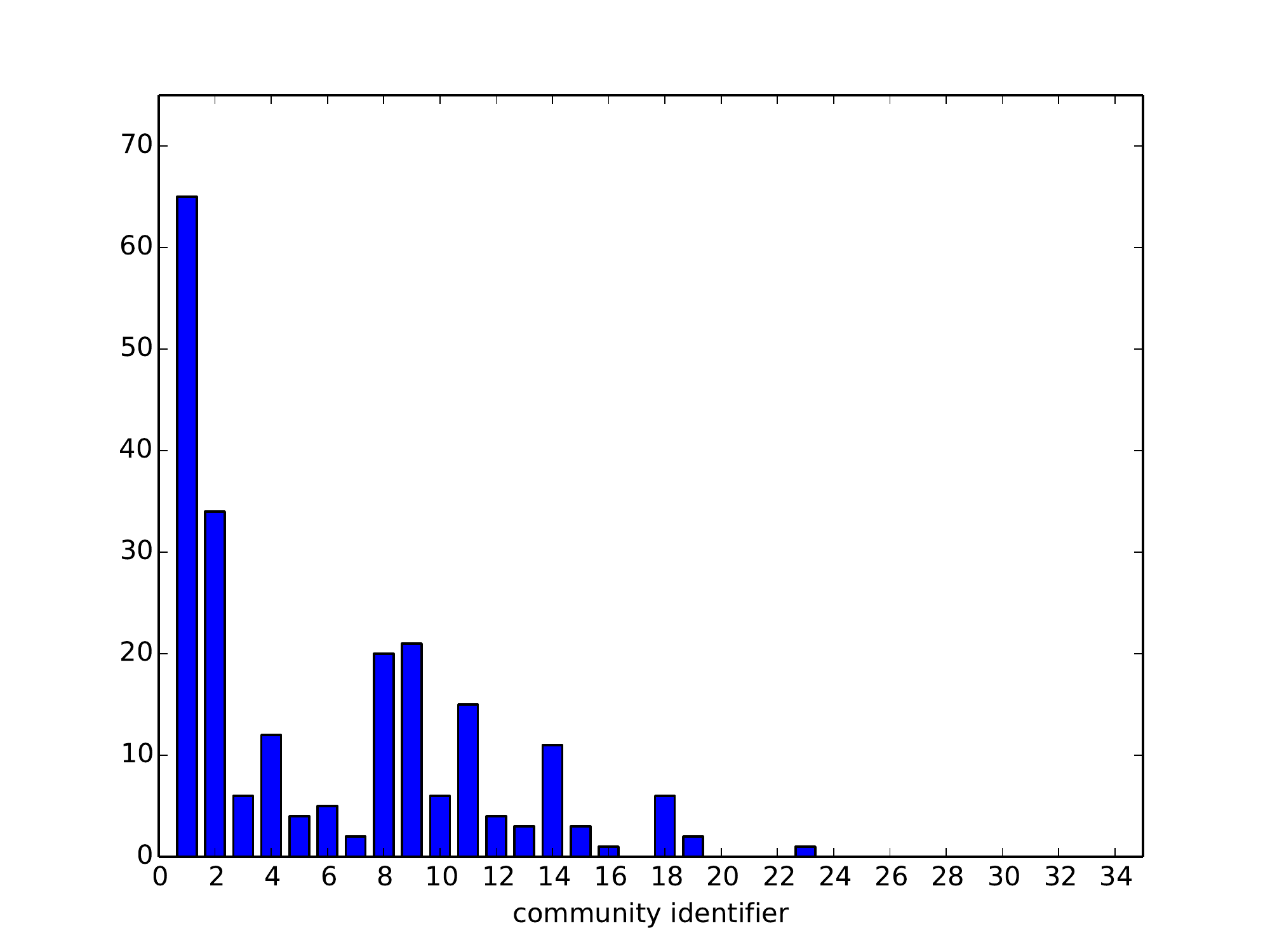}\label{fig:hist692c}}
\subfloat[Forth type, analogous histograms are obtained for $31$ similar nodes.]{\includegraphics[width=.7\columnwidth, angle=0]{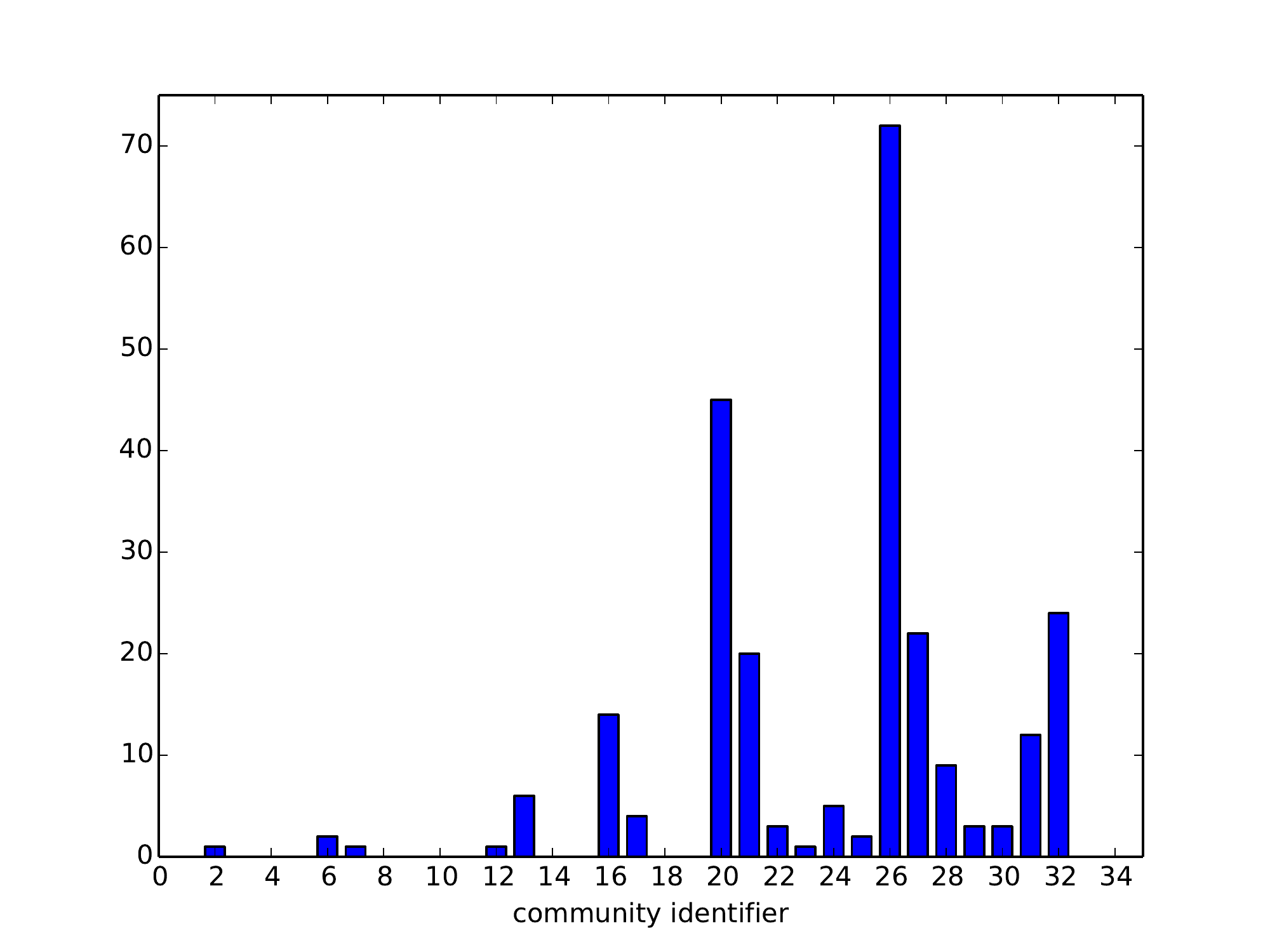}\label{fig:hist692d}}
\caption{Histograms of the frequency of the connection of a given node with original communities of the size different than one for the Koch curve for $N=692$.}
\label{fig:hist692}
\end{center}
\end{figure*}

\subsection{The Sierpinski triangle - the classes}
\begin{figure}[!hptb]
\begin{center}
\includegraphics[width=.7\columnwidth, angle=0]{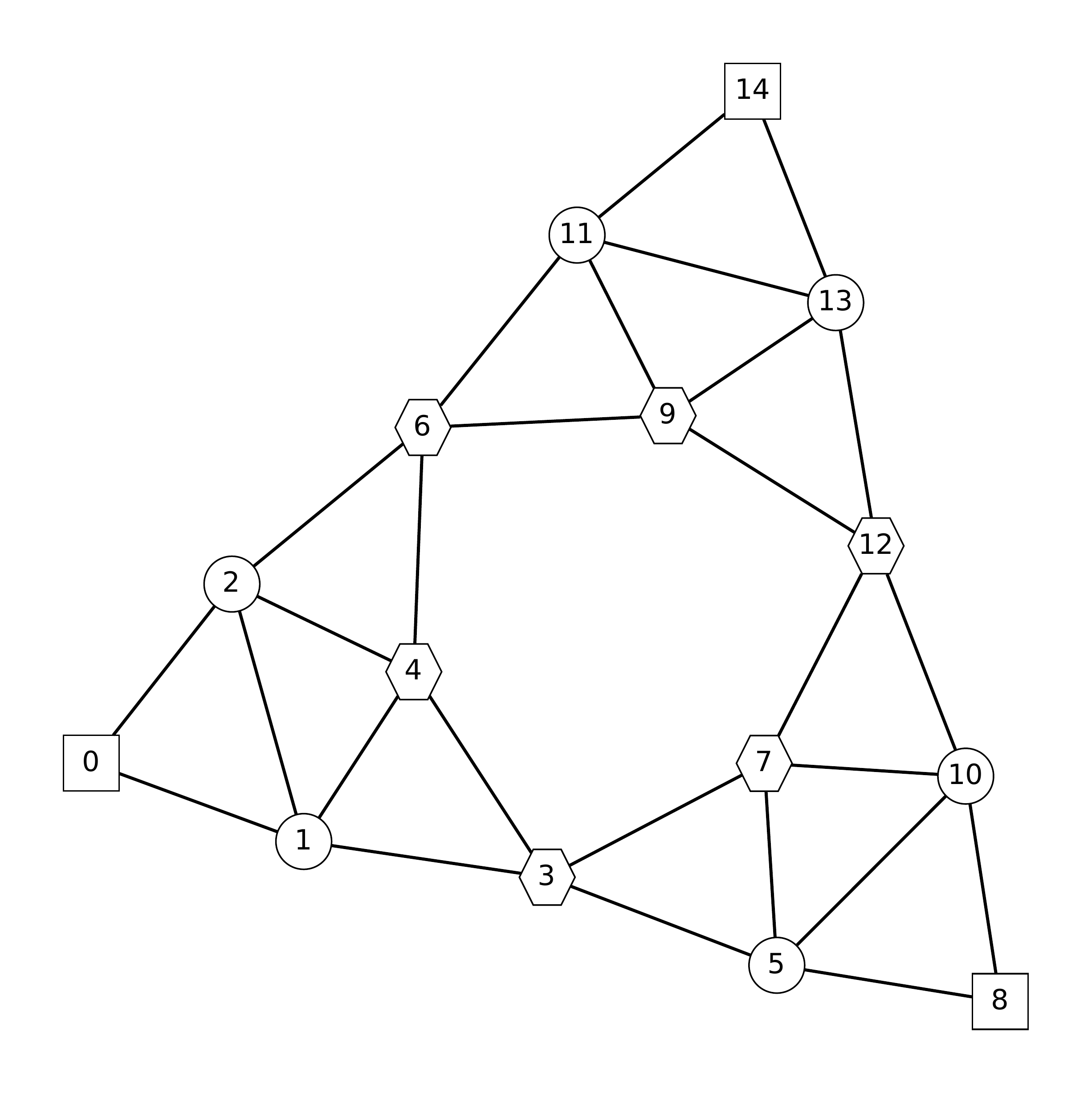}
\caption{The Sierpinski triangle network with indication of nodes classes for $N=15$, here nodes belong to $3$ different classes indicated in the figure by {\large{$\circ$}}, 
{\scriptsize{$\square$}} and {\small{$\hexagon$}}.}
\label{fig:class15}
\end{center}
\end{figure}

The classes obtained for the Sierpinski triangle network of size $N=15$ are presented in Fig.\ref{fig:class15} and Tab.\ref{tab:class15}. Here, nodes are classified to $3$ classes, where the node degree \textit{k} in one class is equal to $2$  and for two remaining classes \textit{k}=$4$.

\begin{table}
\centering
\begin{tabular}{c|c|l}
\textit{id}&\textit{k}&\textit{node's numbers}\\\hline
A&2&0 8 14\\
B&4&1 2 5 10 11 13\\
C&4&3 4 6 7 9 12\\
\end{tabular}
\caption{Classes obtained for the Sierpinski triangle network of size $N=15$. Here nodes are classified to $3$ classes (column \textit{id}), 
for all but one the node degree \textit{k} =$4$.}
\label{tab:class15}
\end{table}

\begin{figure}[!hptb]
\begin{center}
\includegraphics[width=.7\columnwidth, angle=0]{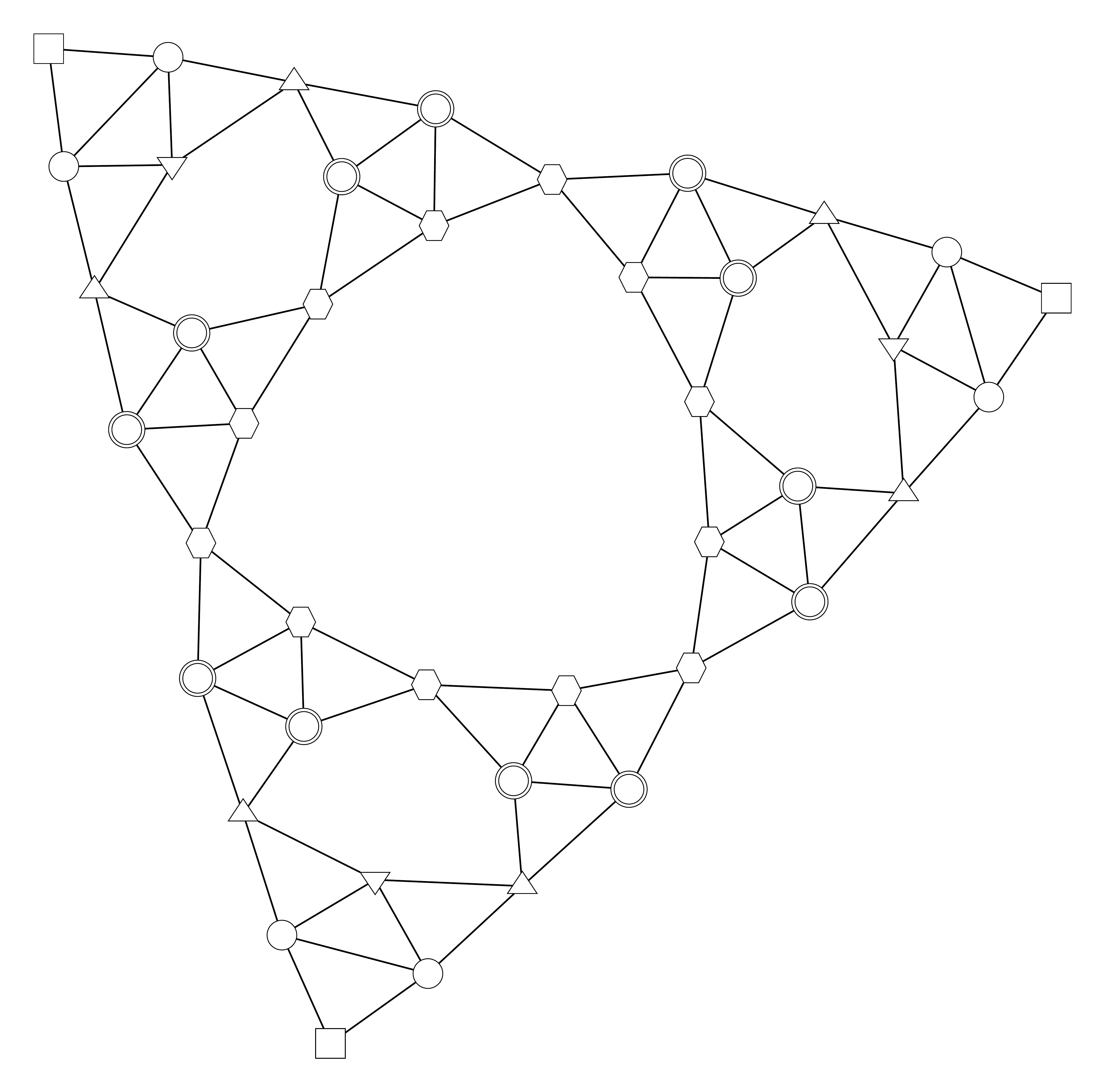}
\caption{The Sierpinski triangle network with indication of nodes classes for $N=42$, here nodes belong to $6$ different classes indicated in the figure by six 
different symbols.}
\label{fig:class42}
\end{center}
\end{figure}

In Fig.\ref{fig:class42} and Tab.\ref{tab:class42} the classes are presented obtained for the network of size $N=42$. In this case $6$ classes are identified, and the node 
degree \textit{k} for all but one (for which $k=2$) is equal to $4$.

\begin{table}
\begin{tabular}{c|c|l}
\textit{id}&\textit{k}&\textit{node's numbers}\\\hline
A&2&0 22 41\\
B&4&4 21 36\\
C&4&1 2 18 25 38 40\\
D&4&3 6 15 29 33 39\\
E&4&5 7 9 11 12 17 24 26 30 32 34 37\\
F&4&8 10 13 14 16 19 20 23 27 28 31 35\\
\end{tabular}
\caption{Classes obtained for the Sierpinski triangle network of size $N=42$. Here nodes are classified to $6$ classes (column \textit{id}), for all but one the 
node degree \textit{k} =$4$.}
\label{tab:class42}
\end{table}

\begin{figure}[!hptb]
\begin{center}
\subfloat[$N=15$]{\includegraphics[width=.3\columnwidth, angle=0]{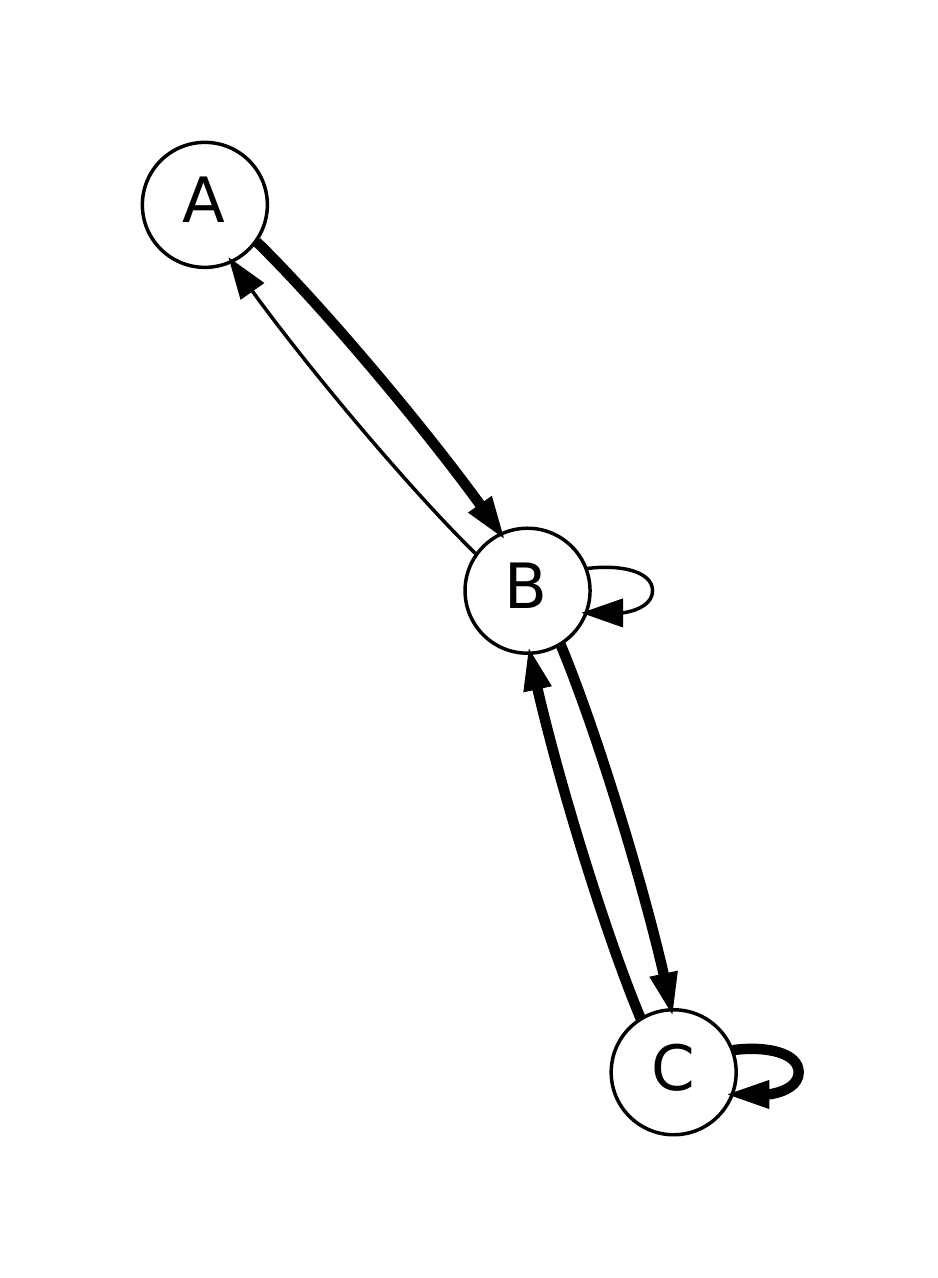}\label{fig:classTSa}}\\
\subfloat[$N=42$]{\includegraphics[width=.4\columnwidth, angle=0]{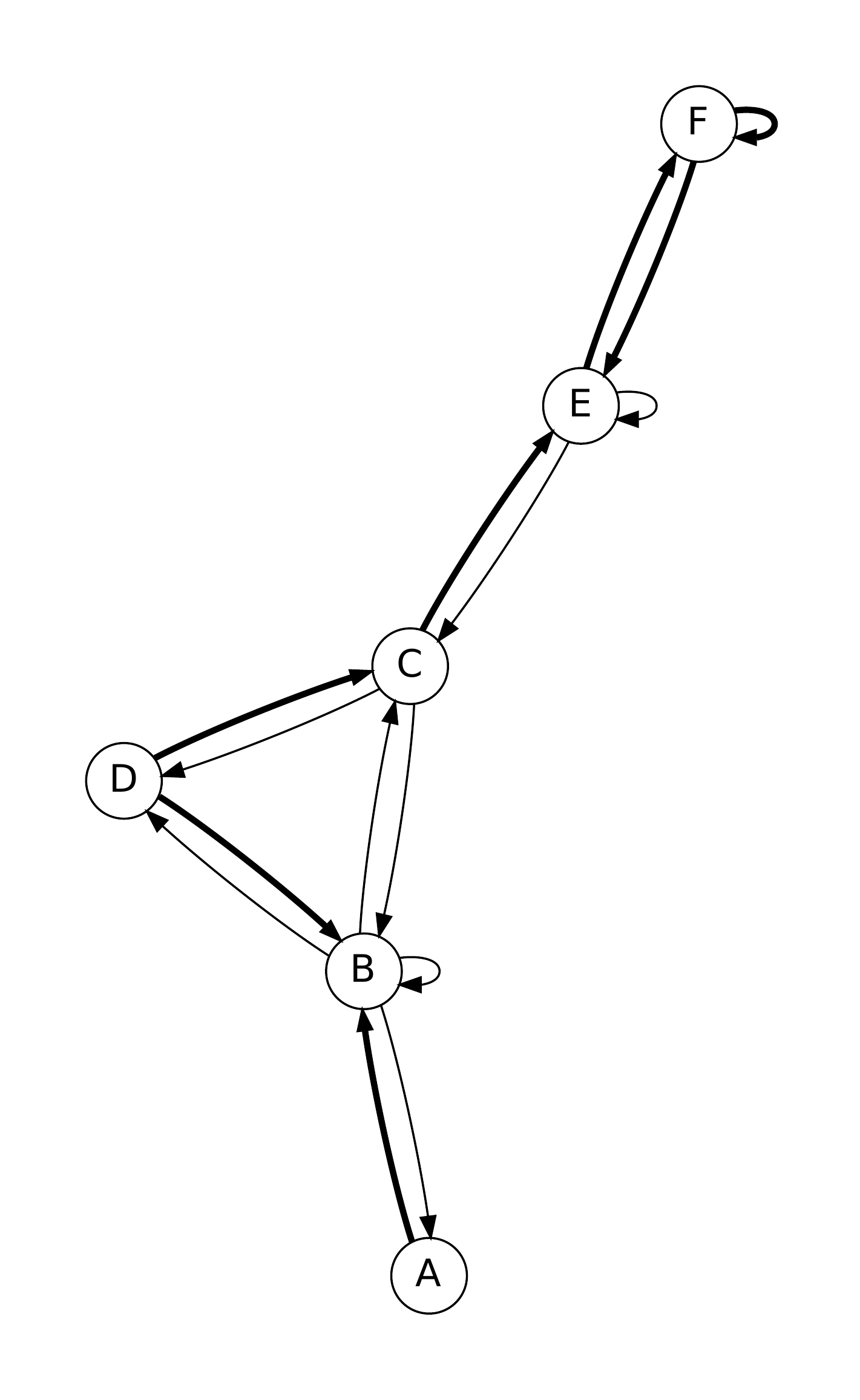}\label{fig:classTSb}}
\caption{Graph of classes for the Sierpinski triangle network.}
\label{fig:classTS}
\end{center}
\end{figure}

Graphs for classes, for both analysed system sizes, are presented in Fig.\ref{fig:classTS}. For example, for $N=15$ there are 3 classes, as there is only three types of nodes. All nodes in class $A$ are connected only to nodes in class $B$. Note that all remaining nodes, although seem topologically different, belong to the same class $C$. This is because each node in $C$ has two neighbours in $C$ and two neighbours in $B$.

\subsection{The Koch curve - the classes}

The procedure of the class identification applied for the Koch curve network indicate to the presence of $5$ classes for $N=116$ (Tab.\ref{tab:class116}) and $17$ classes for $N=692$ (Tab.\ref{tab:class692}). In both cases node degree is equal $2$ or $3$.

\begin{table}
\begin{center}
\begin{tabular}{l*5{|c}}
\textit{id}&A&B&C&D&E\\\hline
\textit{k}&2&2&3&3&3\\\hline
\textit{\#}&12&48&24&24&8
\end{tabular}
\caption{Classes obtained for the Koch curve network of size $N=116$. Here nodes are classified to $5$ classes (row \textit{id}), in $2$ classes the node degree \textit{k}=$2$, and in $3$ classes \textit{k}=$3$, \textit{\#} is the number of nodes belonging to a given class.}
\label{tab:class116}
\end{center}
\end{table}

\begin{table}
\begin{center}
\begin{tabular}{l*9{|c}}
\textit{id}&A&B&C&D&E&F&G&H&I\\\hline
\textit{k}&2&2&2&2&2&2&3&3&3\\\hline
\textit{\#}&12&48&48&96&96&48&24&48&48\\\hline
\textit{id}&J&K&L&M&N&O&P&Q\\\hline
\textit{k}&3&3&3&3&3&3&3&3\\\hline
\textit{\#}&24&48&48&24&24&24&24&8
\end{tabular}
\caption{Classes obtained for the Koch curve network of size $N=692$. Here nodes are classified to $17$ classes (rows \textit{id}), in $6$ classes the node degree \textit{k}=$2$, and in $11$ classes \textit{k}=$3$, \textit{\#} is the number of nodes belonging to a given class.}
\label{tab:class692}
\end{center}
\end{table}

\begin{figure}[!hptb]
\begin{center}
\subfloat[$N=116$]{\includegraphics[width=.25\columnwidth, angle=0]{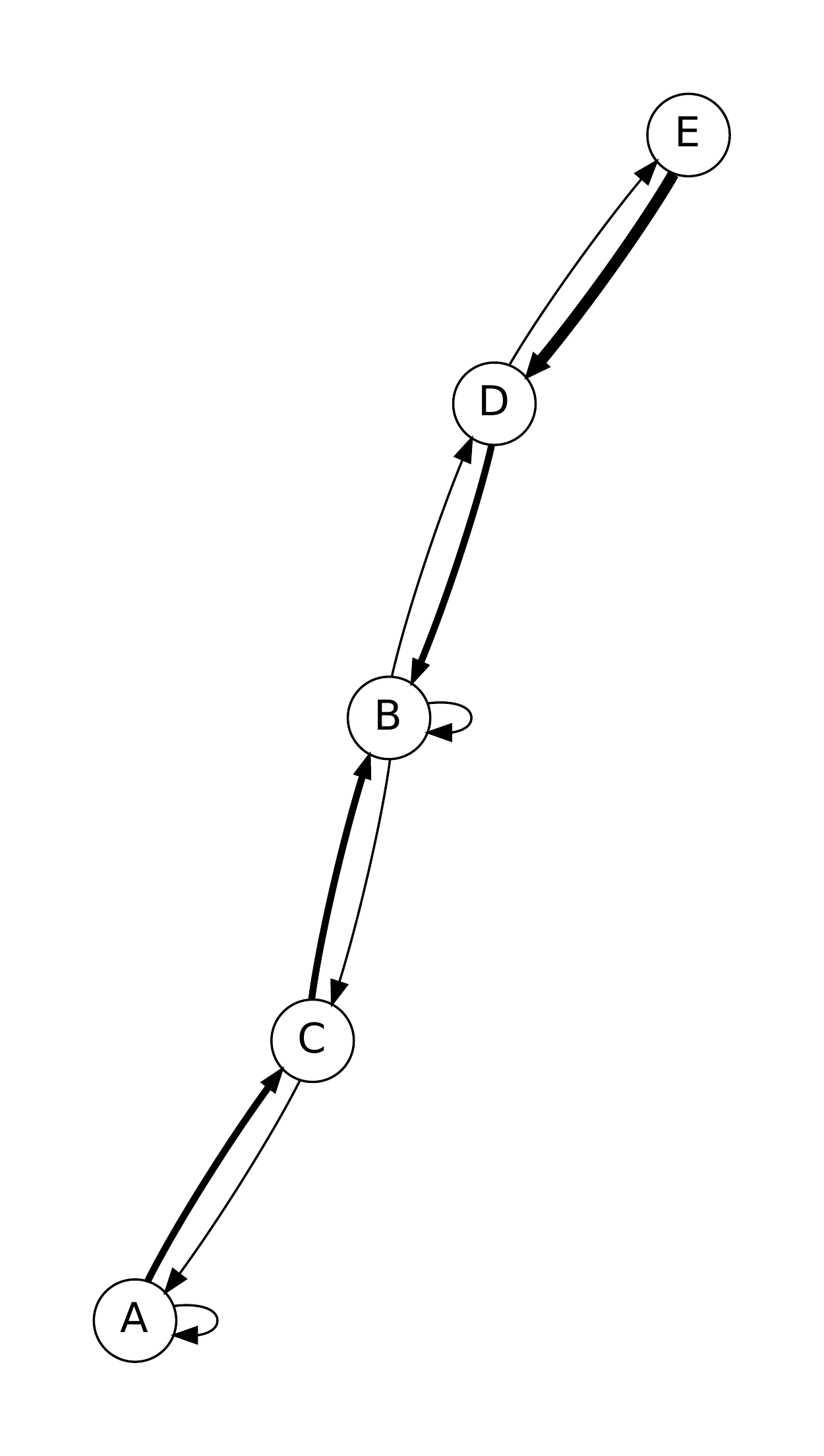}\label{fig:classKCa}}\\
\subfloat[$N=692$]{\includegraphics[width=.9\columnwidth, angle=0]{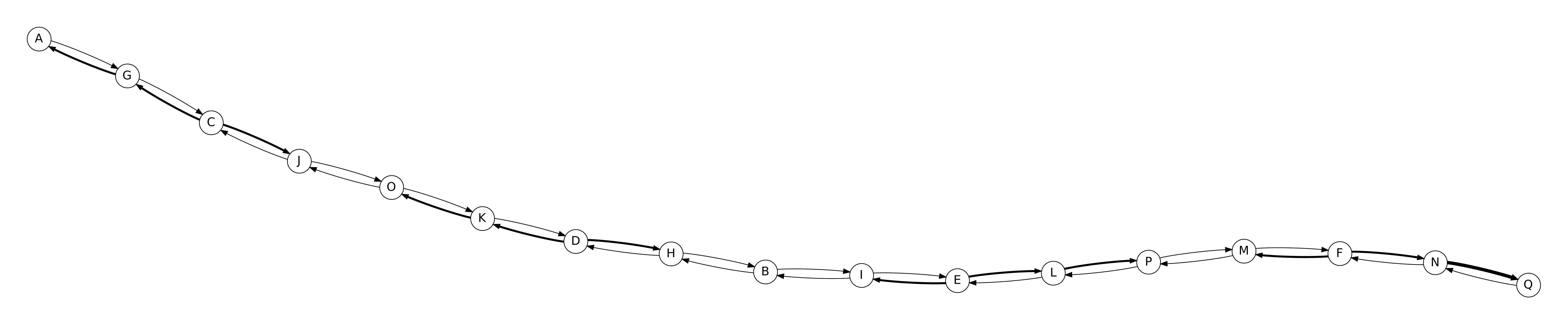}\label{fig:classKCb}}
\caption{Graph of classes for the Koch curve network.}
\label{fig:classKC}
\end{center}
\end{figure}

\subsection{Communities and classes}
Here we ask if the number of classes can be deduced on the basis of knowledge achieved during the communities identification procedure. The analysis of histograms obtained for all 
nodes allows to divide nodes into groups, within which the histogram profiles are similar; actually we believe that with a better statistics, they would be almost the same. This similarity means, that nodes positions in the network are analogous; for example, 
there are nodes which connect two other similar communities. Those similarities also underlie the class identification procedure.\\

Our analysis indicates that nodes which are characterized by the same histograms belong to the same class. In most cases, different histogram profile means that nodes are of different 
classes. However, for example in the case of the Sierpinski triangle network of $N=42$, nodes $8,19, 35$ and $13, 20, 27$ have different histogram profiles, but they belong to the same 
class. 

The analysis of histogram profiles involves only nodes which in the absence of noise form one-node communities, and do not take into account other nodes of the system. 
In some cases they are classified to the classes which cover overlapping nodes, but sometimes they form new classes. 
For small and simple systems, the histogram profiles may be also used to predict the number of types of nodes classified to original communities (with no noise). For each node of the 
same histogram profile one can indicate its nearest neighbours. Then, existence of an intersection 
between lists of neighbours obtained for similar nodes can be checked. Non-zero size of such an intersection means that nodes of different types have a common neighbour.  
And vice versa, if there is no intersection -- nodes are not connected by such a neighborhood. Such information allows us to indicate the number of types of nodes which form communities in the case 
without noise. 

In the case of the Sierpinski triangle of size $N=15$ and $42$ , as it was already written, $2$ and $4$ histogram profiles, respectively, are obtained. We then construct lists each 
of which contains nearest neighbours of nodes for which the same profile is observed. Next, intersection of obtained lists was found. As a result we obtain that in the case of the 
smaller system the lists contain common nodes, while for the larger system intersection of some of lists is an empty set. It appears, that in the first case nodes are of the same type, 
while in the second they can be divided into two categories (see Tab.\ref{tab:interSie}).

\begin{table}
\begin{center}
\begin{tabular}{c*4{|c}}
&\textit{a}&\textit{b}&\textit{c}&\textit{d}\\\hline
\textit{a}&x&x&-&-\\\hline
\textit{b}&x&x&x&x\\\hline
\textit{c}&-&x&x&x\\\hline
\textit{d}&-&x&x&x
\end{tabular}
\caption{Intersection existence between lists of nearest neighbours of nodes for which given histogram profile (\textit{a}, \textit{b}, \textit{c} or \textit{d}) for the Sierpinski triangle network of $N=42$ -- sign 'x' indicate not empty intersection, and '-' empty intersection.}
\label{tab:interSie}
\end{center}
\end{table}

\begin{table}
\begin{center}
\begin{tabular}{c*3{|c}}
&\textit{a}&\textit{b}&\textit{c}\\\hline
\textit{a}&x&-&x\\\hline
\textit{b}&-&x&x\\\hline
\textit{c}&x&x&x
\end{tabular}
\caption{Intersection existence between lists of nearest and next nearest neighbours of nodes for which given histogram profile (\textit{a}, \textit{b} or \textit{c}) for the Koch curve network of $N=116$ -- sign 'x' indicate not empty intersection, and '-' empty intersection.}
\label{tab:interKoch}
\end{center}
\end{table}

In the case of the Koch curve network of size $N=116$, the proper analysis of mentioned intersections involve taking into account also next nearest neighbours. The result obtained both for the Sierpinski triangle network and the Koch curve network are the same as the results obtained from the class identification procedure. For larger systems, or in the case of larger communities such analysis is not sufficient to indicate of nodes classes, as in this case connections schemes of the nodes inside communities are more complicated.

\begin{table*}
\begin{center}
\begin{tabular}{c*7{|c}}
\multirow{2}{*}{\textit{network}}&\multirow{2}{*}{\textit{N}}&\multicolumn{2}{c|}{\textit{no noise}}&\multicolumn{3}{c|}{\textit{with noise}}&\multirow{2}{*}{\textit{\# classes}}\\\cline{3-7}
&&\textit{\# comm.}&\textit{m}&\textit{\# comm.}&\textit{m}&\textit{\# hist.}&\\\hline
\textit{Sierpinski}&15&3(3)+6(1)&0.12&3&0.19&2(1)&3\\\cline{2-8}
\textit{triangle}&42&9(3)+15(1)&0.15&9&0.26&4(2)&6\\\hline
\textit{Koch}&116&24(4)+20(1)&0.32&24&0.42&3(2)&5\\\cline{2-8}
\textit{curve}&692&24(13)+8(40)+60(1)&0.41&32&0.36&4(n.a.)&17\\
\end{tabular}
\caption{Results summary. Abbreviations used in table: \textit{N} - network size, \textit{\# comm.} - number of communities, \textit{m}- modularity, \textit{\# hist.} - number of histogram profiles.}
\label{tab:sum}
\end{center}
\end{table*}

\section{Discussion}

In this paper we present a new application of the formerly proposed system of differential equations \cite{comm} to  identify the structure of overlapping communities. 
The method is applied twice, for unnoised and noised initial values of the coefficients of connectivity matrix. In the first case, part of nodes appear to form one-node communities.
In the second case, these nodes are attached to other communities with different probabilities. The distribution of this probability is found to be a useful characteristics of a 
given node.\\

Next, the class structure of the investigated fractals is determined, according to the method developed in \cite{class,class1}. Having this done, we concentrate on the nodes
which have been found to fall into one-node communities. The result is that once two such nodes are described with the same histograms, they fall into the same class.\\

This procedure may be used for any network, in particular other fractals. \\

{\bf Acknowledgement:}
The author is grateful to Krzysztof~Kułakowski for critical reading of the manuscript and helpful discussions. The work was partially supported by the Polish Ministry of Science and Higher Education and its grants for Scientific Research and by PL-Grid Infrastructure.

\end{document}